\documentclass[12pt]{article}
\usepackage{epsfig}
\def\as{\alpha_{\mbox{\tiny S}}}
\def\VEV#1{\left\langle#1\right\rangle}
\def\tdt{t\frac{\partial}{\partial t}}
\def\lmsb{\Lambda_{\overline{\mbox{\tiny MS}}}}
\def\stot{\sigma_{\mbox{\tiny tot}}}
\newcommand     \lambdamsb     {\ifmmode
          \Lambda^{(5)}_{\overline{\mbox{\tiny MS}}} \else
         $\Lambda^{(5)}_{\overline{\mbox{\tiny MS}}}$ \fi}
\newcommand\addsp[1]{\raisebox{0ex}[1.25\height][1.35\depth]{#1}}
\newcommand\hepex[1]{{\tt hep-ex/}#1}
\newcommand\hepph[1]{{\tt hep-ph/}#1}
\def\cf{{\sl cf.}}
\def\ie{{\sl i.e.}}
\def\etal{{\sl et~al.}}
\def\ibid{{\sl ibid.}}
\def\journal#1;#2,#3(#4){#1{\bf #2},#3(#4)}
\def\Caption#1\endCaption{\caption{#1}}
\def\Eqn#1{Eq.~(\ref{#1})}

\def\Ep#1{(\ref{#1})}
\def\eqnumber{\refstepcounter{equation}\eqno\hbox{(\theequation)}}
\def\EQN #1 {\eqnumber\label{#1}}
\newskip\mycentering
\def\EQNdoublealign#1{
\def\EQN ##1 {&{\refstepcounter{equation}\label{##1}\rm(\theequation)}}
\centerline{\tabskip\mycentering \vbox{\halign to \displaywidth{\hskip\mycentering
                     $\displaystyle\tabskip0pt{##}$                           &
\hskip 2\arraycolsep                    \hfil${##}$\hfil                      &
\hskip 2\arraycolsep $\displaystyle\tabskip0pt{##}$\hfil \tabskip\mycentering & 
                                         \llap{##}\tabskip0pt                  \cr 
                                               #1                              \cr
}}}
\def\EQN ##1 {\eqnumber\label{##1}}
}
\begin{document}
%
\begin{titlepage}
\vspace*{-10mm}
\hbox to \textwidth{ \hsize=\textwidth
\hspace*{0pt\hfill} 
\vbox{ \hsize=58mm
{
\hbox{ Bicocca-FT-01-20 \hss}
\hbox{ Cavendish--HEP--01/12 \hss}
\hbox{ MPI-PhE/2001-14 \hss}
\hbox{ September 28, 2001\hss } 
\hbox{ rev.: November 9, 2001\hss } 
}
}
}

\bigskip\bigskip\bigskip
\begin{center}
{\huge\bf Jet fragmentation in \\[1.5mm]
          e$^{\mathbf{+}}$e$^{\mathbf{-}}$ annihilation
}
\end{center}
\bigskip\bigskip
\begin{center}
{\Large  O.~Biebel$^{(1)}$, 
         P.~Nason$^{(2)}$, 
         B.R.~Webber$^{(3)}$
}
\end{center}
\bigskip
%
\begin{abstract}
\noindent 
A short review of theoretical and experimental results on 
fragmentation in e$^+$e$^-$ annihilation is presented. Starting
with an introduction of the concept of fragmentation functions 
in e$^+$e$^-$ annihilation, aspects of scaling violation, 
multiplicities, small and large $x$, longitudinal, gluon, 
light and heavy quark fragmentation are summarized. 

\end{abstract}

\vspace*{0pt\vfill}
\vfill
\bigskip

{
\small
\noindent
$^{(1)}$ 
\begin{minipage}[t]{160mm} 
Max-Planck-Institut f\"ur Physik, D-80805 M\"unchen, Germany \\
contact e-mail: biebel@mppmu.mpg.de 
\end{minipage}\\
$^{(2)}$ 
\begin{minipage}[t]{160mm} 
INFN, Sez.\ di Milano, Piazza della Scienza 3, 20126 Milano, Italy \\
contact e-mail: paolo.nason@mi.infn.it
\end{minipage}\\
$^{(3)}$ 
\begin{minipage}[t]{160mm} 
Cavendish Laboratory, University of Cambridge, Madingley Road, Cambridge, CB3 0HE, UK \\
contact e-mail: webber@hep.phy.cam.ac.uk
\end{minipage}\\
}
\end{titlepage}
\newpage
\tableofcontents
\newpage
\section{Introduction}
Fragmentation functions are dimensionless functions that
describe the final-state single-particle energy distributions
in hard scattering processes. The total $e^+e^-$ fragmentation function
for hadrons of type $h$ in annihilation at c.m.\ energy
$\sqrt{s}$, via an intermediate vector boson $V=\gamma/Z^0$, is
defined as
$$
F^h(x,s) = \frac{1}{\stot}\frac{d\sigma}{dx}
(e^+e^-\to V\to h X)
\EQN Fxs $$
where $x=2E_h/\sqrt{s}\le 1$ is the scaled hadron
energy (in practice, the approximation
$x=x_p=2p_h/\sqrt{s}$ is often used). Since each hadron of
type $h$ contributes to the fragmentation function \Ep{Fxs},
its integral with respect to $x$ gives the average multiplicity of 
those hadrons:
$$
\VEV{n_h(s)} = \int_0^1 dx\,F^h(x,s)\;.
\EQN avnint $$
Similarly, the integral of $x\,F^h(x,s)$ gives the total scaled
energy carried by hadrons of that type, which implies the sum
rule
$$
\sum_h\int_0^1 dx\,x\,F^h(x,s) = 2\;.
\EQN momint $$

   Neglecting contributions suppressed by inverse powers of $s$,
the fragmentation function \Ep{Fxs} can be represented as
a sum of contributions from the different parton types
   $i=u, \bar{u}, d,\bar{d}, \ldots ,g$:
$$
F^h(x,s) = \sum_i\int_x^1 \frac{dz}{z} C_i(s;z,\as) D^h_i(x/z,s)\;.
\EQN Fsum $$
where $D_i^h$ are the parton fragmentation functions.
In lowest order the coefficient function $C_g$ for gluons is zero,
while for quarks $C_i=g_i(s)\delta(1-z)$ where $g_i(s)$ is the
appropriate electroweak coupling.  In particular, $g_i(s)$ is
proportional to the charge-squared of parton $i$ at $s\ll M_Z^2$,
when weak effects can be neglected. 
   In higher orders the coefficient functions and parton fragmentation 
   functions are factorization-scheme dependent.

Parton fragmentation functions are analogous to the parton distributions
in deep inelastic scattering (Section~9 of \cite{bib-RPP00}). 
In the fragmentation function $x$ represents the fraction of
a parton's momentum carried by a produced hadron, whereas
in a parton distribution it represents the fraction of a
hadron's momentum carried by a constituent parton.
In both cases, the simplest parton-model approach
would predict a scale-independent $x$ distribution. 
Furthermore we obtain similar violations of this scaling
behaviour when QCD corrections are taken into account.

\section{Scaling violation}\label{sec:jetscaviol}
The evolution of the parton fragmentation function $D_i(x,t)$ with
increasing scale $t=s$, like that of the parton distribution function
$f_i(x,t)$ with $t=s$ (see Section~35 of \cite{bib-RPP00}), 
is governed 
by the DGLAP equation \cite{DGLAP}
$$
\tdt D_i(x,t)
= \sum_j\int_x^1\frac{dz}{z} \frac{\as}{2\pi}
P_{ji}(z,\as) D_j(x/z,t) \;.
\EQN APDi $$
   In analogy to DIS, in some cases an evolution equation for the 
   fragmentation function $F$ itself (\Eqn{Fsum}) can be derived 
   from \Eqn{APDi} \cite{NW}.
Notice that the splitting function is now $P_{ji}$ rather
than $P_{ij}$ since here $D_j$ represents the fragmentation
of the final parton, whereas $f_j$ in 
Eqs.~(35.7) - (37.10) of \cite{bib-RPP00}  
represented the distribution of the initial parton. The
splitting functions again have perturbative expansions of the 
form 
$$
P_{ji}(z,\as) =P^{(0)}_{ji}(z)
+\frac{\as}{2\pi} P^{(1)}_{ji}(z) +\cdots
\EQN APkernels $$
where the lowest-order functions $P^{(0)}_{ji}(z)$
are the same as those in deep inelastic scattering
but the higher-order terms \cite{bib-Furmanski-PLB97-437}\footnote{There 
are misprints in the formulae in the published article. The 
correct expressions can be found in the preprint version or
in \cite{bib-ESW-book}.} are different. The effect
of evolution is, however, the same in both cases: as the
scale increases, one observes a scaling violation in
which the $x$ distribution is shifted towards lower values.
This can be seen from Fig.~\ref{fig-inclfrag}.
\figure 
\centerline{\epsfysize=12cm \epsfbox{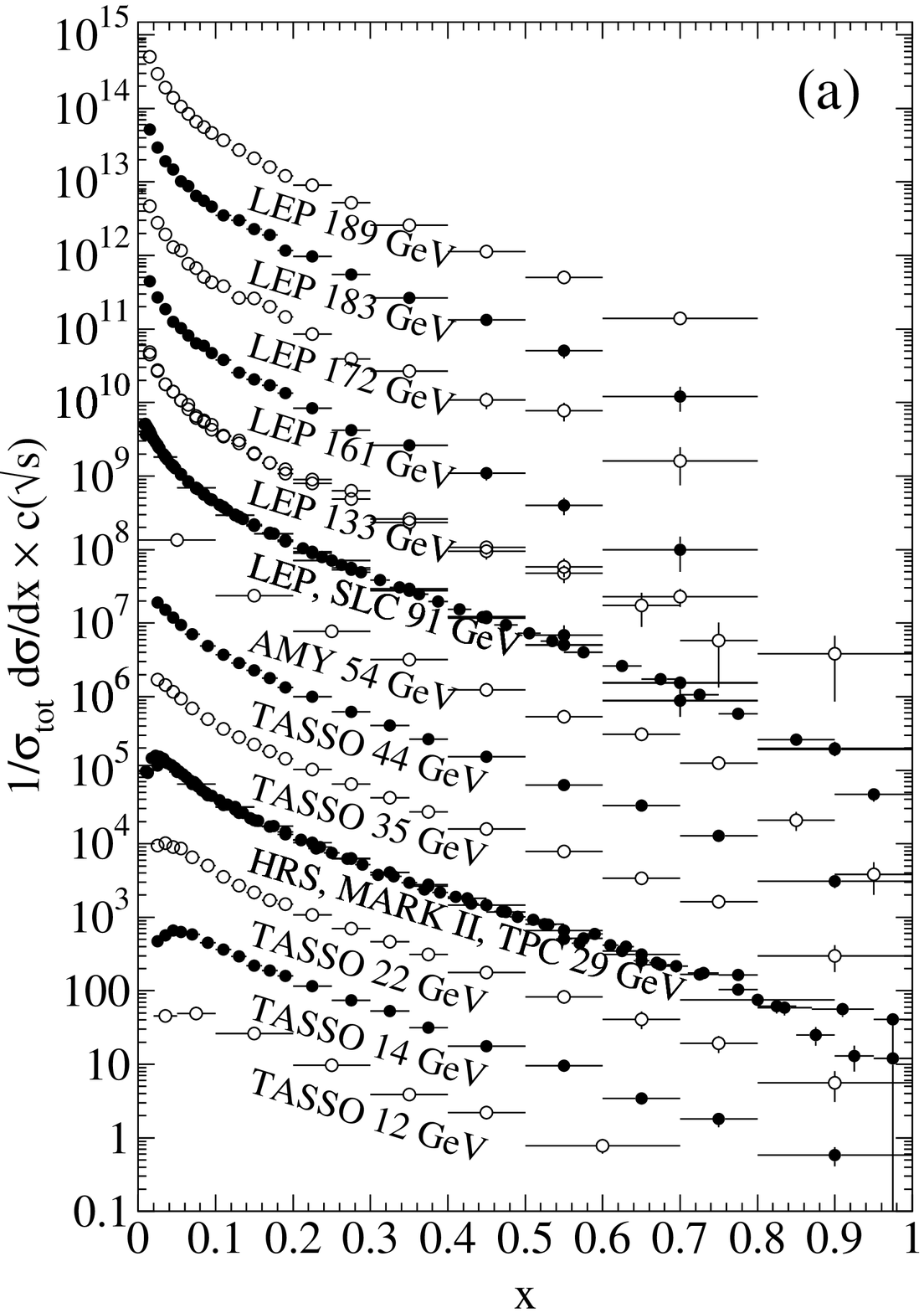}
            \epsfysize=10cm \epsfbox{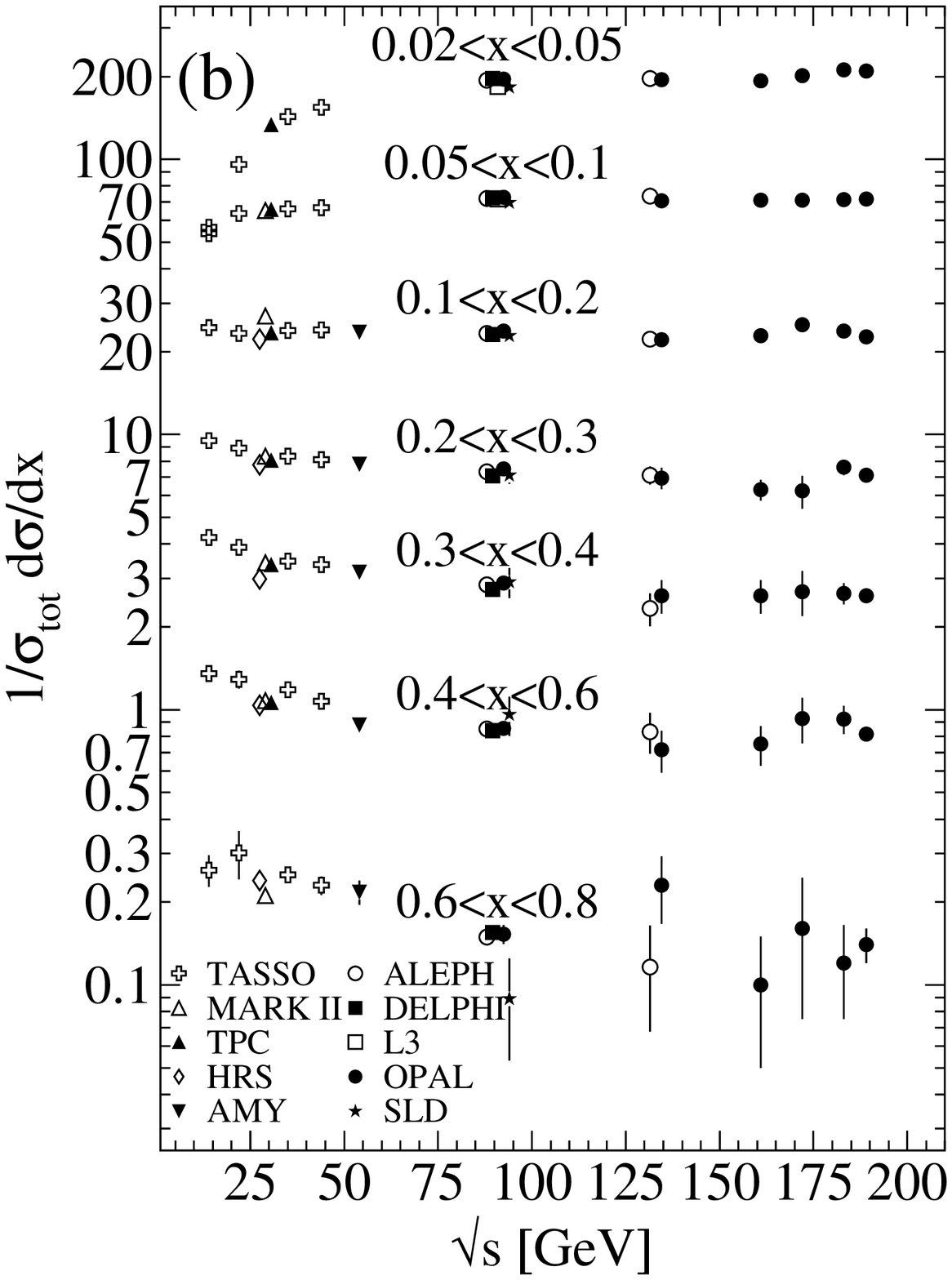}}
\Caption
         \label{fig-inclfrag}
         The $e^+e^-$ fragmentation function for all charged particles 
         is shown (a) for different c.m.\ energies, $\sqrt{s}$, 
         versus $x$ and (b) for various ranges of $x$ versus $\sqrt{s}$.
         For the purpose of plotting (a), a scale factor $c(\sqrt{s})=10^i$
         was applied to the distributions where $i$ is ranging from $i=0$ 
        ($\sqrt{s}=12$~GeV) to $i=12$ ($\sqrt{s}=189$~GeV).
         Data are compiled from references
         \cite{ bib-TASSO-PLB114-65, 
                bib-TASSO-ZPC47-187, 
                bib-HRS-PRD31-1,
                bib-MARKII-PRD37-1, 
                bib-TPC-PRL61-1263, 
                bib-AMY-PRD41-2675,
                bib-ALEPH-PREP294-1, 
                bib-DELPHI-EPJC6-19, 
                bib-L3-PLB259-199, 
                bib-OPAL-EPJC7-369, 
                bib-MARKII-PRL64-1334, 
                bib-ALEPH-ZPC73-409,
                bib-OPAL-ZPC72-191, 
                bib-OPAL-ZPC75-193, 
                bib-OPAL-EPJC16-185}.  
\endCaption
\endfigure

As in the case of the parton distribution functions,
the most common strategy for solving the evolution
equations \Ep{APDi} is to take moments
(Mellin transforms) with respect to $x$:
$$
\tilde D(j,t) = \int_0^1 dx\; x^{j-1}\; D(x,t)\; ,
\EQN Mellin $$
the inverse Mellin transformation being
$$
D(x,t) = \frac{1}{2\pi i}\int_C dj\; x^{-j}\; \tilde D(j,t)\;,
\EQN invMell $$
where the integration contour $C$ in the complex $j$ plane
is parallel to the imaginary axis and to the right of all
singularities of the integrand.
Again we can consider fragmentation function
combinations which are non-singlet (in flavour space) such as
$D_V=D_{q_i} - D_{\bar q_i}$ or $D_{q_i} -D_{q_j}$. In these
combinations the mixing with the flavour singlet gluons drops
out and for a fixed value of $\as$ the solution is simply
$$
\tilde D_V(j,t) = \tilde D_V(j,t_0)
\left(\frac{t}{t_0}\right)^{\gamma_{qq}(j,\as)}  \;,
\EQN DGLAPsol $$
where the quantity $\gamma_{qq}$ is the Mellin
transform of $\as P_{qq}/2\pi$.
The resulting dependence on $t$ violates scaling rules
based on naive dimensional analysis, and therefore the
power $\gamma_{qq}$ is called an {\it anomalous dimension}.
For a running coupling $\as(t)$, however, the scaling
violation is no longer power-behaved. Inserting the
lowest-order term in Eq.~(9.5a) of \cite{bib-RPP00}   
for the running coupling,
we find that the solution varies like a power of $\ln t$:
$$
\EQNdoublealign{
\tilde D_V(j,t) &=& \tilde D_V(j,t_0) \left(\frac{\as
(t_0)}{\as (t)}\right)^{d_{qq}(j)} 	, \nonumber \cr
d_{qq}(j) &=& \frac{6C_F}{11C_A-2n_f}\left[-\frac 1 2
+ \frac{1}{j(j+1)} - 2\sum_{k=2}^j\frac 1 k \right]\; ,
\EQN scaviol %
}
$$
where $C_F=4/3$, $C_A=3$, and $n_f$ is the number of quark
flavours as in Section~9 of \cite{bib-RPP00}.   

For the singlet fragmentation function
$$
D_S=\sum_i (D_{q_i} + D_{\bar q_i})
\EQN DSdef $$
we have mixing with the fragmentation of the gluon and the
evolution equation becomes a matrix relation
as in the deep inelastic scattering case:
$$
t\frac{\partial}{\partial t}
 \left(\begin{array}{c} \tilde D_S \cr \tilde D_g \end{array}\right)
= \left(\begin{array}{cc} \gamma_{qq} & 2n_f\gamma_{gq} \cr
\gamma_{qg} & \gamma_{gg} \end{array}\right)
\left(\begin{array}{c} \tilde D_S \cr \tilde D_g \end{array}\right)\; .
\EQN singev $$
The anomalous dimension matrix in this equation has two
real eigenvalues,
$$
\gamma_{\pm}={1\over 2}[\gamma_{gg}+\gamma_{qq}\pm
\sqrt{(\gamma_{gg}-\gamma_{qq})^2+8n_f\gamma_{gq}\gamma_{qg}}]\;.
\EQN gammaEigenval $$
Expressing $D_S$ and $D_g$ as linear combinations of the
corresponding eigenvectors $D_+$ and $D_-$, we find that they
evolve as superpositions of terms of the form \Ep{scaviol}
with $\gamma_+$ and $\gamma_-$ in the place of $\gamma_{qq}$.

At small $x$, corresponding to $j\to 1$, the
$g\to g$ anomalous dimension becomes dominant and we find
$\gamma_+\to \gamma_{gg}\to\infty$, $\gamma_-\to\gamma_{qq}\to 0$.
This region requires special treatment, which will be presented
in Section~\ref{sec_smallxfrag}.

There are several complications in the experimental study of
scaling violation in jet fragmentation functions \cite{NW}.
First, the energy dependence of the electroweak couplings
$g_i(s)$ that enter into \Eqn{Fsum}
is especially strong in the energy region presently under study
($\sqrt{s} =20-200$ GeV).  In particular, the $b$-quark contribution
varies by more than a factor of 2 in this range.
The fragmentation of the $b$ quark into
charged hadrons, including the weak decay products of the
$b$-flavoured hadron, is expected to be substantially softer
than that of the other quarks, so its varying contribution
can give rise to a `fake' scaling violation.
A smaller, partially compensating effect is
expected in charm fragmentation. These effects can be
eliminated by extracting the $b$ and $c$ fragmentation
functions from tagged heavy quark events. 
Fig.~\ref{fig-flavdepfrag} shows the flavour-dependent $e^+e^-$ 
fragmentation functions for quarks determined at $\sqrt{s}=91$ GeV.

Secondly, one requires the gluon fragmentation function $D_g(x,s)$
in addition to those of the quarks.  Although the gluon does not
couple directly to the electroweak current, it contributes
in higher order, and mixes with the quarks through evolution.
Its fragmentation can be studied in tagged heavy-quark
three-jet ($Q\bar Q g$) events, or via the longitudinal
fragmentation function (to be discussed below).
Results for both methods are shown in Fig.~\ref{fig-flavdepfrag}.
\figure 
\centerline{\epsfysize=8cm \epsfbox{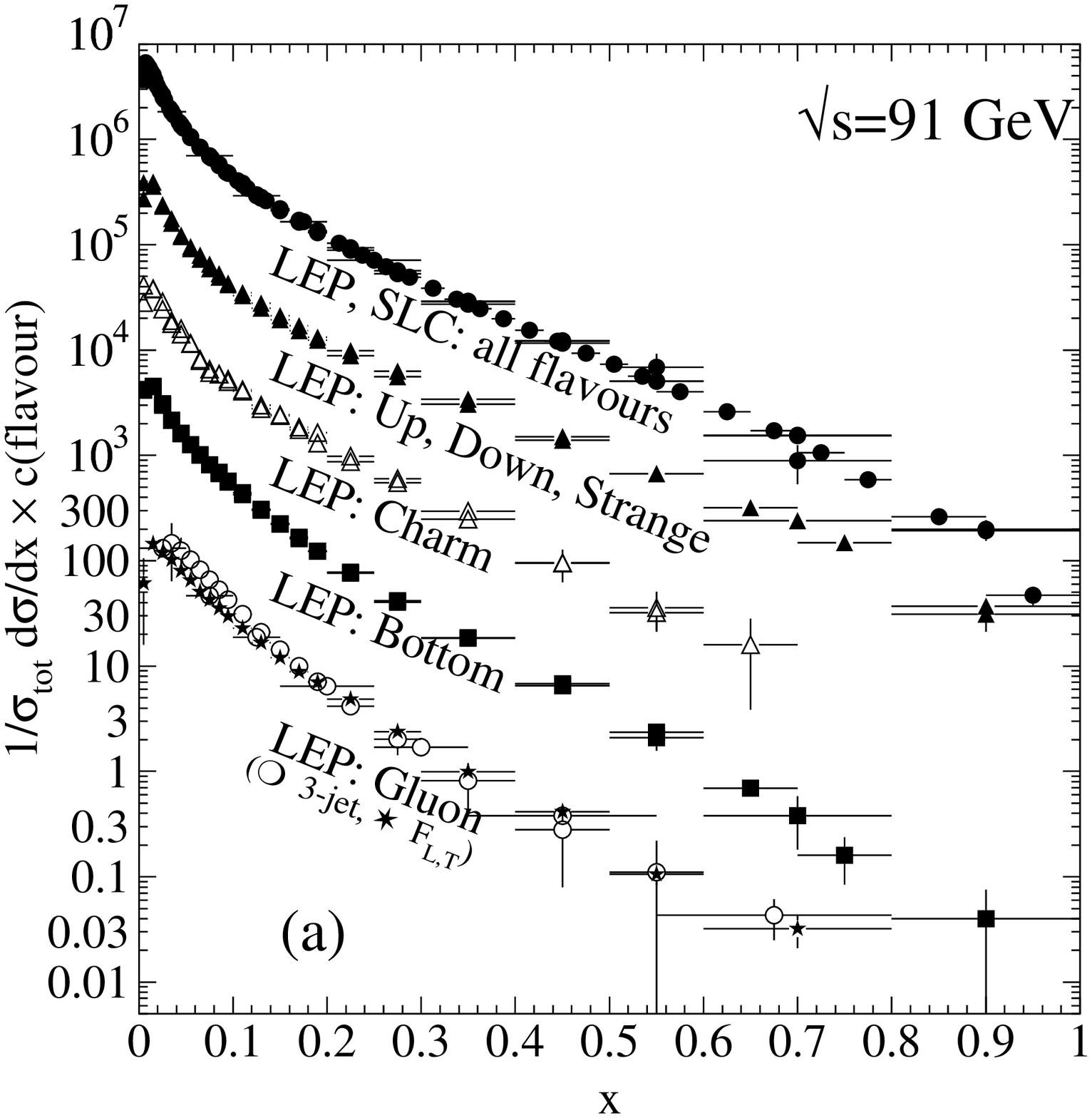} 
            \epsfysize=8cm \epsfbox{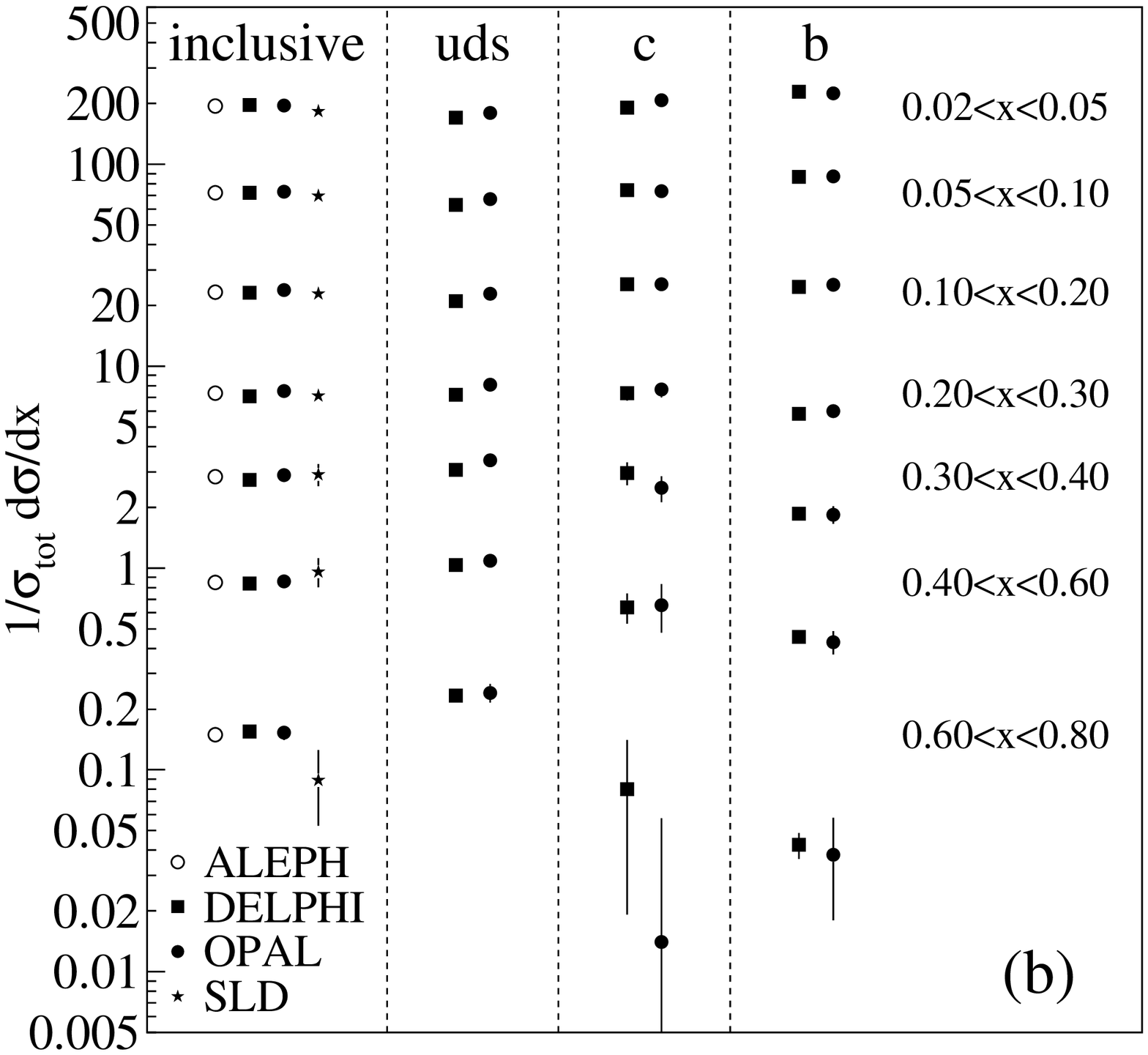}}
\Caption
        \label{fig-flavdepfrag}
         Comparison of the charged-particle and the 
         flavour-dependent $e^+e^-$ fragmentation functions 
         obtained at $\sqrt{s}=91$~GeV. The data are shown (a) for 
         the inclusive, light (up, down, strange) quarks, 
         charm quark, bottom quark, and the gluon versus $x$,
         and (b) in various ranges of $x$. 
         For the purpose of plotting (a), a scale factor $c({\rm flavour})=10^i$
         was applied to the distributions where $i$ is ranging from $i=0$ 
         (Gluon) to $i=4$ (all flavours).
         Data are compiled from references 
         \cite{bib-ALEPH-PREP294-1, 
               bib-DELPHI-EPJC6-19,
               bib-L3-PLB259-199, 
               bib-OPAL-EPJC7-369, 
               bib-MARKII-PRL64-1334,
               bib-ALEPH-EPJC17-1, 
               bib-DELPHI-PLB398-194,
               bib-OPAL-EPJC11-217, 
               bib-OPAL-ZPC68-203,
               bib-OPAL-ZPC68-179}.
\endCaption
\endfigure

A final complication is that power corrections to fragmentation
functions, of the form $f(x)/Q^p$, are not so well understood
as those in deep inelastic scattering.  Theoretical arguments
\cite{BalBra, DasWeb} suggest that hadronization can lead
to $1/s$ corrections. Therefore, possible contributions of this
form should be included in the parametrization when fitting the
scaling violation.
More conservatively, one could also try to fit $1/Q$
corrections to check for their presence.

Quantitative results of studies of scaling violation in $e^+e^-$
fragmentation are reported in refs.~\cite{bib-DELPHI-PLB398-194, 
                                          bib-DELPHI-PLB311-408, 
                                          bib-ALEPH-PLB357-487, 
                                          bib-DELPHI-EPJC13-573,
                                          bib-Kniehl-PRL85-5288}.
The values of $\as$ obtained are consistent with the world average
in Section~9 of \cite{bib-RPP00}.     

\section{Average Multiplicities}\label{sec_average_multiplicity}
The average number of hadrons of type $h$ in the fragmentation
of a parton of type $i$ at scale $t$,
$\VEV{n_h(t)}_i$, is just the integral of the fragmentation
function, which is the $j=1$ moment in the notation of
\Eqn{Mellin}:
$$
\VEV{n_h(t)}_i = \int_0^1 dx\, D^h_i(x,t) = \tilde D^h_i(1,t)\; .
\EQN eqn-mean_mult $$
If we try to compute the $t$ dependence of this quantity, we
immediately encounter the problem that the lowest-order
expressions for the anomalous dimensions $\gamma_{gq}$ and
$\gamma_{gg}$ in \Eqn{singev} are divergent at $j=1$.
The reason is that for $j \le 1$ the moments of the
splitting functions in \Eqn{APkernels} are dominated by the
region of small $z$, where $P_{gi}(z)$ has a divergence
associated with soft gluon emission.

In fact, however, we can still solve the evolution equation
for the average multiplicity provided we take into account
the suppression of soft gluon emission due to coherence 
\cite{bib-colour-coherence,book}.
The leading effect of coherence is that the scale on the
right-hand side of the DGLAP equation \Eqn{APDi} is reduced
by a factor of $z^2$:
$$
\tdt D_i(x,t) = \sum_j\int_x^1\frac{dz}{z} \frac{\as}{2\pi}
P_{ji}(z,\as) D_j(x/z,z^2 t)\;.
\EQN APtlt $$
This change is not important for most values of $x$ but it is
crucial at small $x$. The anomalous dimensions are now found to 
be finite at $j=1$ and
\Eqn{eqn-mean_mult} gives to next-to-leading order 
\cite{Mueller}:
$$
\VEV{n(s)} = a\cdot \exp\left[
\frac{4}{\beta_0}\sqrt{\frac{6\pi}{\as(s)}}
+\left(\frac{1}{4}+\frac{10n_f}{27\beta_0}
\right)\ln\as(s)\right]
+c\;.
\EQN avmult_ho $$
where $a$ and $c$ are constants, and $\beta_0$ is defined in
Eq.~(9.4b) of \cite{bib-RPP00}.    
The resulting prediction, shown by the curve in
Fig.~{\ref{fig_avmult}} after fitting the parameters $a$ and
$c$,
is in very good agreement with experiment.
\figure 
\centerline{\epsfxsize=10cm \epsfbox{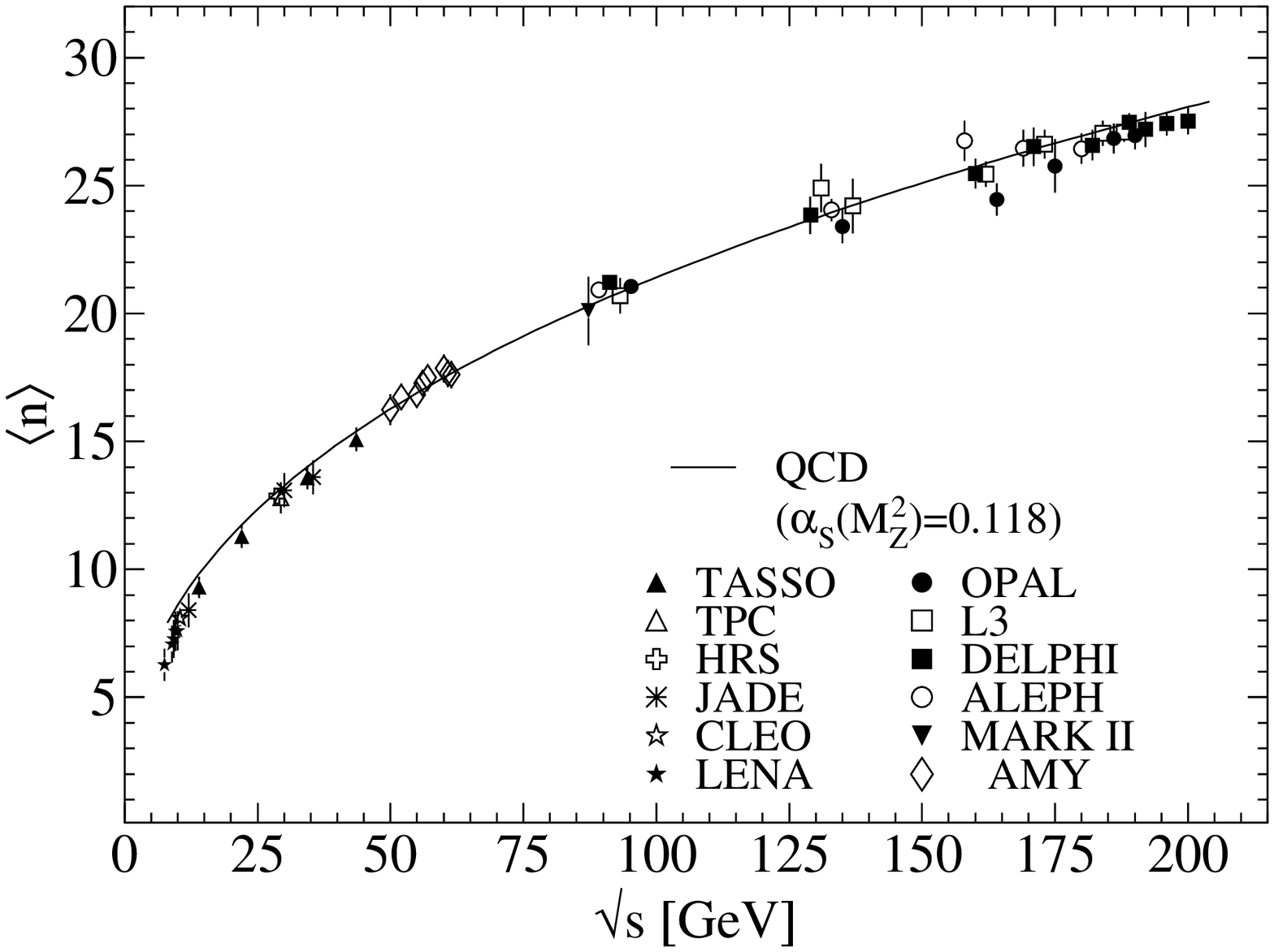}}
\Caption
         \label{fig_avmult}
         The mean multiplicity of charged particles measured in
         $e^+e^-$ annihilation by various experiments is shown
         \cite{bib-ALEPH-PREP294-1,
               bib-DELPHI-EPJC6-19,
               bib-L3-PLB259-199,
               bib-MARKII-PRL64-1334,
               bib-ALEPH-ZPC73-409,
               bib-OPAL-ZPC72-191,
               bib-OPAL-ZPC75-193,
               bib-OPAL-EPJC16-185,
               bib-OPAL-ZPC68-203,
               bib-LENA-ZPC9-1,
               bib-CLEO-PRL49-357,
               bib-JADE-ZPC20-187,
               bib-HRS-PRD34-3304,
               bib-TPC-PL134B-299,
               bib-AMY-PRD42-737,
               bib-ALEPH-98-025,
               bib-DELPHI-PLB372-172,
               bib-DELPHI-PLB416-233,
               bib-L3-PLB371-137,
               bib-L3-PLB404-390,
               bib-L3-2304,
               bib-L3-98-148}.
         The measurements include contributions from $K^0_S$
         and $\Lambda$ decays. Overlaid is the prediction \Eqn{avmult_ho}
         using $\as(M_Z^2)=0.118$. 
\endCaption
\endfigure
In this comparison the scale parameter $\Lambda_{\mbox{\scriptsize mult}}$ 
in $\as(s)$ was held fixed.
$\Lambda_{\mbox{\scriptsize mult}}$ is not necessarily equal to $\lmsb$,
because the renormalization scheme dependence
of $\VEV{n(s)}$ only appears at next-to-next-to-leading
order, and corrections to \Eqn{avmult_ho} of this
order have not yet been calculated.  In fact, however,
the value used in Fig.~{\ref{fig_avmult}},
$\Lambda_{\mbox{\scriptsize mult}}=226$ MeV,
corresponding to $\as(M_Z^2)=0.118$, is 
close to $\lmsb$, indicating that further higher-order
corrections should be small. Higher order corrections to 
\Eqn{avmult_ho} have been considered in
\cite{bib-Dremin, bib-Dremin-Nechitailo,
bib-Dremin-Gary-PLB459-341}.

\section{Small-$x$ Fragmentation}
\label{sec_smallxfrag}
\index{fragmentation function>small-$x$}
The behaviour of $\tilde D(j,s)$, \Eqn{Mellin}, near
$j=1$ determines the form of small-$x$
fragmentation functions via the inverse Mellin transformation
\Ep{invMell}. Keeping the first three terms in a Taylor
expansion of the anomalous dimension $\gamma_{gg}$ 
around $j=1$ gives a simple Gaussian function of $j$
which transforms into a Gaussian in the variable
$\xi\equiv\ln(1/x)$:
$$
xD(x,s)\propto \exp\left[-\frac{1}{2\sigma^2}(\xi-
\xi_p)^2\right]\;,
\EQN xigauss $$
where the peak position is
$$
\xi_p = \frac{1}{4b\as(s)}\simeq
\frac{1}{4}\ln\left(\frac{s}{\Lambda^2}\right)
\EQN xipeak $$
and the width of the distribution of $\xi$ is
$$
\sigma = \left(\frac{1}{24b}\sqrt{\frac{2\pi}{C_A\as^3(s)}}
\right)^{\frac{1}{2}} \propto
\left[\ln\left(\frac{s}{\Lambda^2}\right)\right]^{\frac{3}{4}}
\;.
\EQN xiwidth $$
Again, one can compute next-to-leading corrections to these
predictions. In the method of \cite{book}, the
corrections are included in an analytical form known as
the `modified leading logarithmic approximation'
(MLLA). Alternatively \cite{FongWeb} they can be used to
compute the higher moment corrections to the Gaussian form
\Ep{xigauss}.
Fig.~\ref{fig-xifrag} shows the $\xi$ distribution at
various c.m.\ energies.
\figure 
\centerline{\epsfxsize=10cm \epsfbox{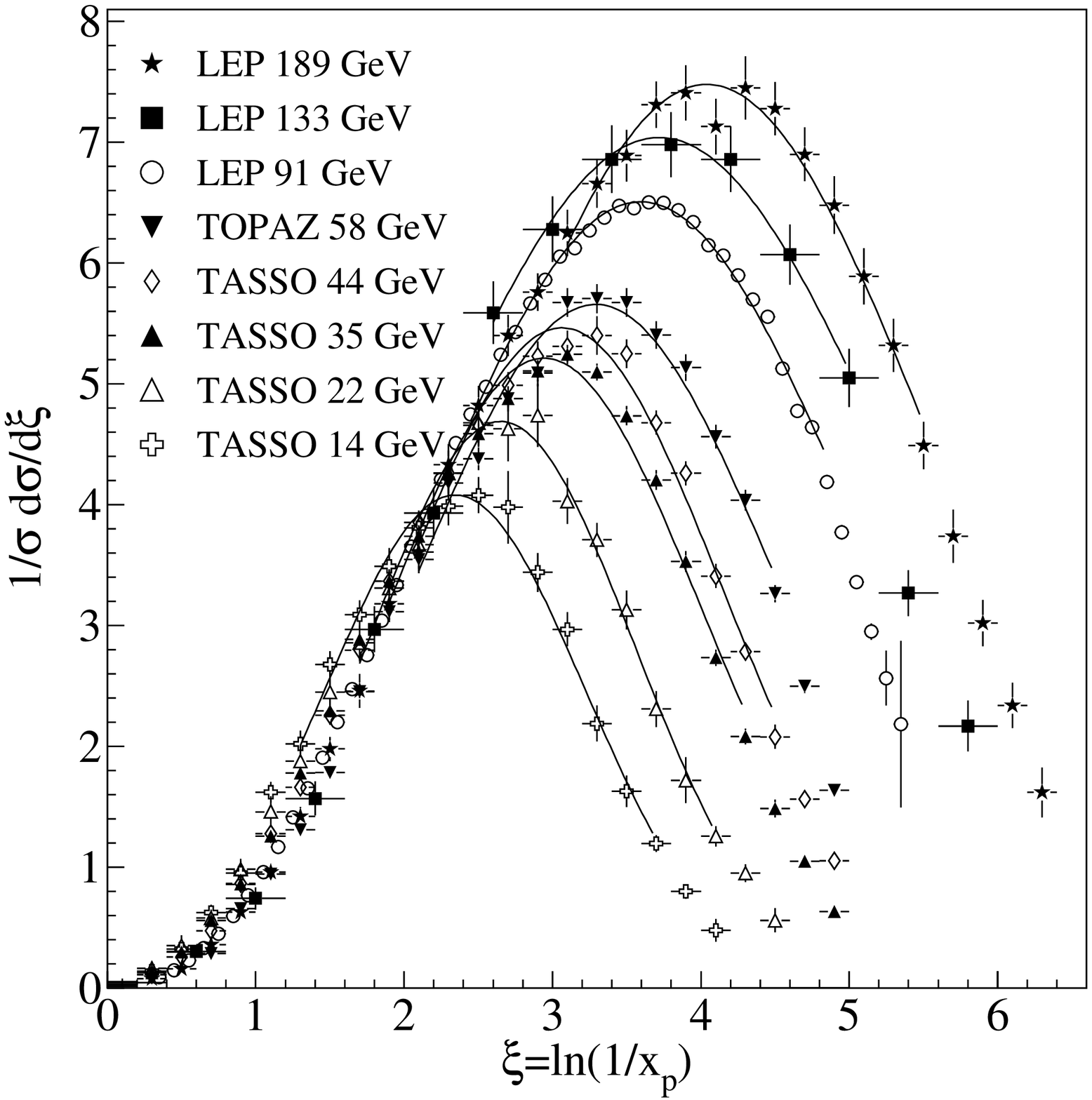}}
\Caption
         \label{fig-xifrag}
         Distribution of $\xi=\ln(1/x_p)$ at several c.m.\ energies 
         \cite{bib-TASSO-ZPC47-187, 
                        bib-ALEPH-PREP294-1, 
                        bib-L3-PLB259-199,
                        bib-OPAL-EPJC7-369, 
                        bib-ALEPH-ZPC73-409, 
                        bib-OPAL-ZPC72-191, 
                        bib-OPAL-ZPC75-193, 
                        bib-OPAL-EPJC16-185,
                        bib-TOPAZ-PLB345-335,
                        bib-DELPHI-ZPC73-11, 
                        bib-DELPHI-ZPC73-229}.
         At each energy only one representative measurement
         is shown and measurements between 133 and 189 GeV are not
         shown due to their small statistical significance. Overlaid
         are fits of a simple Gaussian function for illustration.
\endCaption
\endfigure   

The predicted energy dependence \Ep{xipeak} of the peak
in the $\xi$ distribution is a striking illustration of soft gluon
coherence, which is the origin of the suppression of hadron production
at small $x$. Of course, a decrease at very small $x$ is
expected on purely kinematical grounds, but this would occur
at particle energies proportional to their masses, i.e.\ at
$x\propto m/\sqrt{s}$ and hence $\xi\sim{1\over 2}\ln s$. 
Thus if the suppression were purely kinematic the peak
position $\xi_p$ would vary twice as rapidly with energy,
which is ruled out by the data (see Fig.~{\ref{fig_peak}}).
\figure 
\centerline{\epsfxsize=10cm \epsfbox{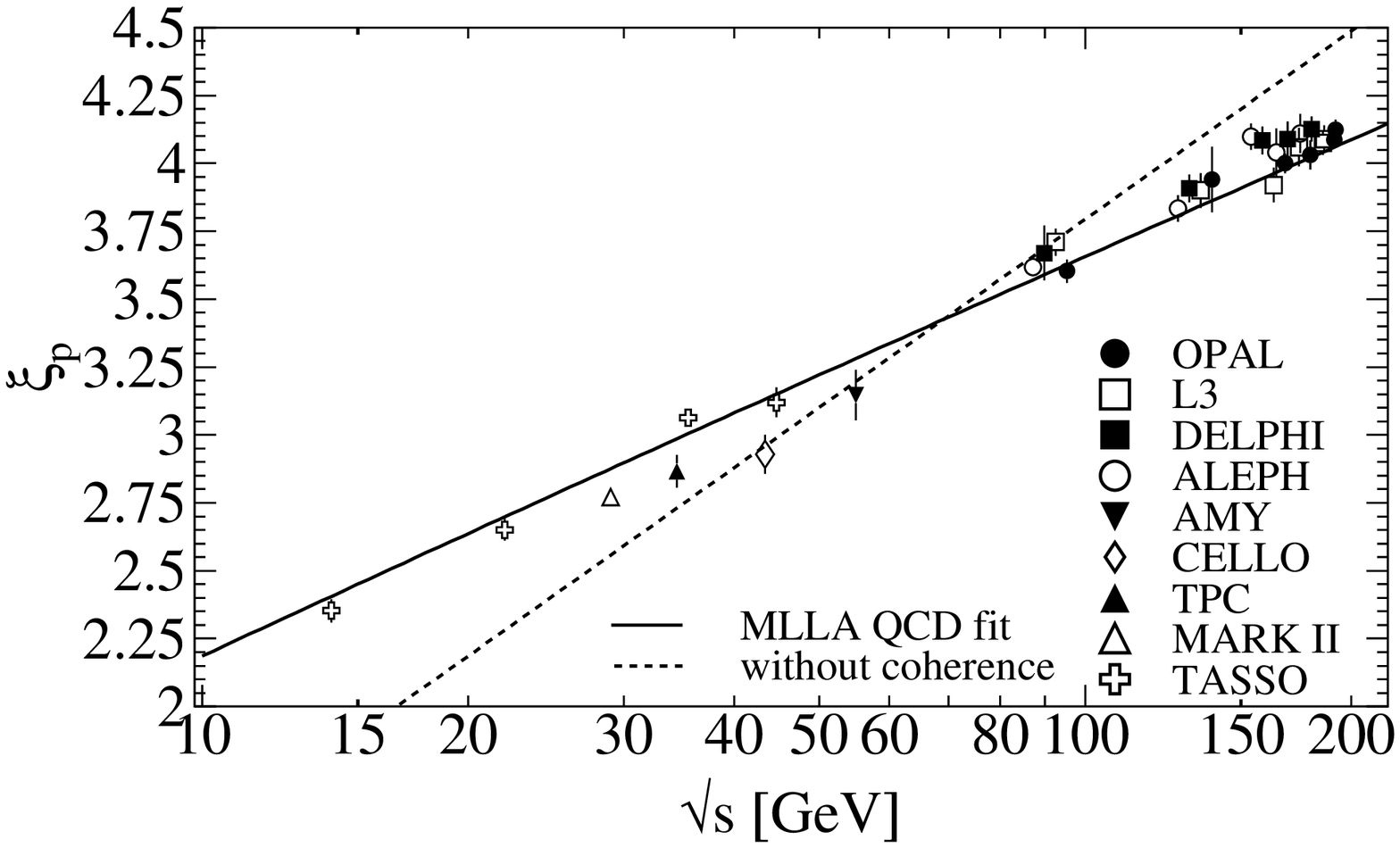}}
\Caption
         \label{fig_peak}
         Evolution of the peak position, $\xi_p$, of the $\xi$
         distribution with the c.m.\ energy $\sqrt{s}$.
         The MLLA QCD prediction (solid) and the expectation without 
         gluon coherence (dashed) were fitted to the data 
         \cite{bib-TASSO-ZPC47-187,
               bib-MARKII-PRD37-1,
               bib-AMY-PRD41-2675,
               bib-L3-PLB259-199,
               bib-ALEPH-ZPC73-409,
               bib-OPAL-ZPC72-191,
               bib-OPAL-ZPC75-193,
               bib-OPAL-EPJC16-185,
               bib-ALEPH-98-025,
               bib-L3-2304,
               bib-L3-98-148,
               bib-DELPHI-ZPC73-229,
               bib-TPC-LBL-23737,
               bib-ALEPH-ZPC55-209,
               bib-DELPHI-PLB275-231,
               bib-DELPHI-98-83,
               bib-OPAL-PLB247-617}.
\endCaption
\endfigure

\section{Large-$x$ Fragmentation}
At large values of the energy fraction $x$, there are enhancements
in the coefficient functions $C_i$ in \Ep{Fsum} and the splitting
functions $P_{ji}$ in \Eqn{APDi}.  These are associated with the
emission of soft and/or collinear gluons, which can lead to factors
of $\as^n\ln^m(1-x)/(1-x)$ with $m\leq 2n-1$ in the $n$th order of
perturbation theory.  It turns out that in the conventional
$\overline{\mbox{MS}}$ factorization scheme the largest terms,
with $m>n$, all occur in the coefficient functions. After the Mellin
transformation \Ep{Mellin}, they are resummed by the
following expression \cite{CatTre,CacCat}
$$
\ln\tilde C_q(j,s) = -\int_0^1 dz\frac{z^{j-1}-1}{1-z}
\left\{\int_{(1-z)s}^s\frac{dt}{t} A[\as(t)] + B[\as((1-z)s]\right\}
\EQN large-x $$
where $A(\as)$ and  $B(\as)$ have perturbative expansions
$$
A(\as) = \sum_{n=1}^\infty \left(\frac{\as}{\pi}\right)^n A^{(n)}\;,\;\;\;
B(\as) = \sum_{n=1}^\infty \left(\frac{\as}{\pi}\right)^n B^{(n)}
\EQN A+B-def $$
with
$$
\EQNdoublealign{
A^{(1)} &=& C_F = \frac 4 3\;,                                             \cr
A^{(2)} &=& C_F\left[C_A\left(\frac{67}{36}-\frac{\pi^2}{12}\right)
-\frac{5}{18}n_f\right] = \frac{67}{9}-\frac{\pi^2}{3}-\frac{10}{27}n_f\;, \EQN A1A2B1-def \cr
B^{(1)} &=& \frac{3}{4}C_F = 1\;.  
}
$$

\section{Longitudinal Fragmentation}
In the process $e^+e^-\to V\to h X$, the joint distribution in the
energy fraction $x$ and the angle $\theta$ between the observed
hadron $h$ and the incoming electron beam has the general form
$$
\frac{1}{\stot}\frac{d^2\sigma}{dx\,d\cos\theta}
= \frac 3 8 (1+\cos^2\theta)\,F_T(x)
+ \frac 3 4    \sin^2\theta \,F_L(x)
+ \frac 3 4    \cos  \theta \,F_A(x)\; ,
\EQN sigTLA $$
where $F_T$, $F_L$ and $F_A$ are respectively the
transverse, longitudinal and asymmetric
fragmentation functions. All these functions also depend
on the c.m.\ energy $\sqrt{s}$.
   \Eqn{sigTLA} is the most general form of the inclusive single
   particle production from the decay of a massive vector 
   boson \cite{NW}.
As their names imply, $F_T$ and $F_L$ represent the
contributions from virtual bosons polarized transversely or
longitudinally with respect to the direction of motion of the
hadron $h$.  $F_A$ is a parity-violating contribution
which comes from the interference between the 
   vector and axial vector
contributions.  Integrating over all angles, we obtain the
total fragmentation function, $F=F_T +F_L$.
Each of these functions can be represented as a convolution
of the parton fragmentation functions $D_i$ with appropriate
coefficient functions $C_i^{\mbox{\scriptsize T,L,A}}$
as in \Eqn{Fsum}. 
   This representation works in the high energy limit. Furthermore,
   when $x\cdot\sqrt{s}$ is of the order of a few hundred MeV,
   the $p_\perp$ smearing of the parton can no longer be neglected, 
   and the fragmentation function formalism no longer account correctly 
   for the separation of $F_T$, $F_L$, and $F_A$.
The transverse and longitudinal coefficient functions are \cite{CFP,AEMP}
$$
\EQNdoublealign{
C_q^{\mbox{\scriptsize T}}(z) &=& \delta(1-z) + O(\as) \nonumber \cr
C_g^{\mbox{\scriptsize T}}(z) &=& O(\as) \nonumber \cr
C_q^{\mbox{\scriptsize L}}(z) &=& C_F\frac{\as}{2\pi} + O(\as^2) \nonumber \cr
C_g^{\mbox{\scriptsize L}}(z) &=& 4C_F\frac{\as}{2\pi}\left(\frac{1}{z}
-1\right) + O(\as^2)\;,
\EQN TLcoeff %
}
$$
which implies that
$$
F_L(x) = C_F\frac{\as}{2\pi}\int_x^1\frac{dz}{z}
\Bigl[ F_T(z)
+ 4\left(\frac z x -1\right)D_g(z)\Bigr] + O(\as^2)\;.
\EQN FL $$
Thus the gluon fragmentation function $D_g$ can be extracted
from measurements of $F_T$ and $F_L$. The next-to-leading order
corrections to \Eqn{FL} are also known \cite{RvN,Binn}.
In Fig.~\ref{fig-ltfrag} $F_T$, $F_L$, and $F_A$ measured at $\sqrt{s}=91$ GeV
are shown. The gluon fragmentation function derived from $F_T$ and $F_L$ in 
\cite{bib-OPAL-ZPC68-203} is shown in Fig.~\ref{fig-flavdepfrag}(a).
\figure 
\centerline{\epsfysize=14cm \epsfbox{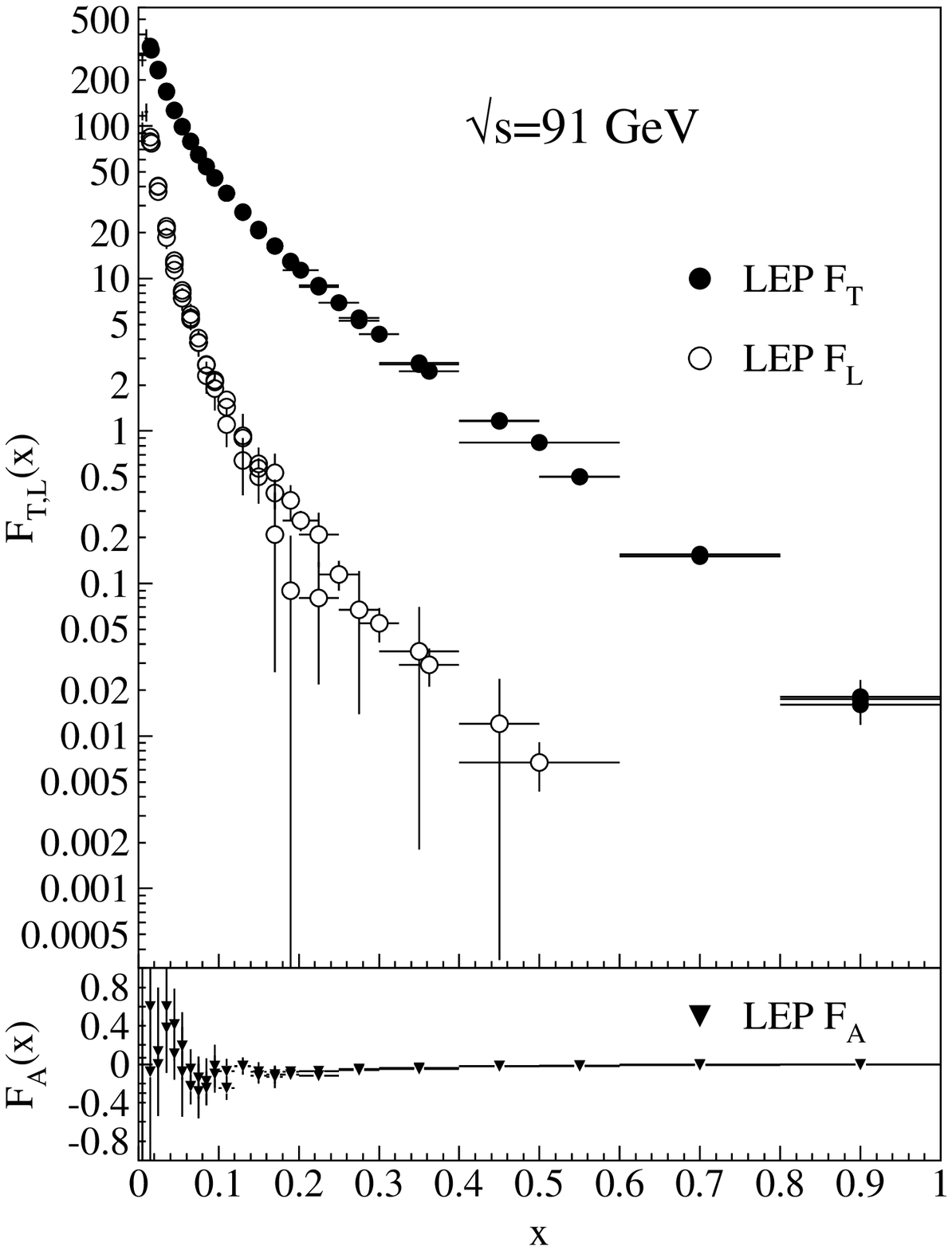}}
\Caption
         \label{fig-ltfrag}
         Transverse ($F_T$), longitudinal ($F_L$), and asymmetric ($F_A$) 
         fragmentation functions are shown 
         \cite{bib-DELPHI-EPJC6-19, 
               bib-OPAL-ZPC68-203,
               bib-ALEPH-PLB357-487}. 
        Data points with
         relative errors greater than $100\%$ are not shown.
\endCaption
\endfigure

Summed over all particle types, the total fragmentation function
satisfies the energy sum rule \Ep{momint}, which we may write as
$$
\frac 1 2\int_0^1 dx\,xF(x) = 1\;.
\EQN sumrule $$
Similarly the integrals
$$
\frac 1 2\int_0^1 dx\,xF_{T,L}(x) \equiv
\frac{\sigma_{T,L}}{\stot}
\EQN sigma_T,L $$
give the transverse and longitudinal fractions of the total
cross section.  The perturbative prediction is \cite{RvN} 
$$
\EQNdoublealign{
\frac{\sigma_{L}}{\stot} &=& \frac{\as}{\pi} +
\left(\frac{601}{40}-\frac{6}{5}\zeta(3)-\frac{37}{36}n_f\right)
\left(\frac{\as}{\pi}\right)^2+ O(\as^3)                           \cr
&\simeq&  \frac{\as}{\pi} +
(13.583-1.028n_f)\left(\frac{\as}{\pi}\right)^2
\EQN sigl %
}
$$
where $\zeta(3)=1.202\ldots$.
Comparing with Eq.~(9.17) of \cite{bib-RPP00},  
we see that the whole 
of the $O(\as)$ correction to $\stot$ comes from the longitudinal 
part, while the $O(\as^2)$ correction receives both longitudinal and
transverse contributions.

\section{Gluon fragmentation}
As we saw in the previous Section, the gluon fragmentation function
can be extracted from the longitudinal fragmentation function using
\Eqn{FL}. It can also be deduced from the fragmentation of
three-jet events in which the gluon jet is identified, for example
by tagging the other two jets with heavy quark decays.  The trouble
with the latter method is that the relevant coefficient function has
not been computed to next-to-leading order, and so the scale and scheme
dependence of the extracted fragmentation function is ambiguous.
The experimentally measured gluon fragmentation functions are shown in
Fig.~\ref{fig-flavdepfrag}(a).

\section{Fragmentation models}
Although the scaling violation can be calculated perturbatively,
the actual form of the parton fragmentation functions is 
non-perturbative. Perturbative evolution gives rise to a
shower of quarks and gluons (partons).
Phenomenological schemes are then used to model the carry-over of 
parton momenta and flavour to the hadrons. Two
of the very popular models are the {\it string fragmentation}
\cite{bib-Artru-Mennessier,bib-Andersson-Gustafson}, implemented 
in the JETSET \cite{bib-JETSET} and UCLA \cite{bib-UCLA} Monte Carlo 
event generation programs, 
and the {\it cluster fragmentation} of the HERWIG Monte Carlo
event generator \cite{Marchesini92}. 

\subsection{String fragmentation}
The string-fragmentation scheme considers the colour field between
the partons, \ie, quarks and gluons, 
   to be the fragmenting entity rather than the partons themselves. 
   The string can be viewed as a colour flux tube formed by 
gluon self-interaction as two coloured
partons move apart. Energetic gluon emission is regarded as 
energy-momentum carrying ``kinks'' on the string. When the energy 
stored in the string is sufficient, a $q\bar{q}$ pair may be created 
from the vacuum. Thus the string breaks up repeatedly into colour 
singlet systems as long as the invariant mass of the string pieces
exceeds the on-shell mass of a hadron. The $q\bar{q}$ pairs are 
created according to the probability of a tunneling process 
       $\exp(-\pi m_{q,\perp}^2/\kappa)$ 
which depends on the transverse mass squared 
       $m_{q,\perp}^2 \equiv m_q^2+p_{q,\perp}^2$ 
and the string tension $\kappa\approx 1$ GeV/fm. The transverse momentum 
$p_{q,\perp}$ is locally compensated between quark and antiquark. Due to 
the dependence on the 
   parton mass $m_q$ and/or the hadron mass $m_h$, 
the production of strange and, in particular, heavy-quark hadrons is suppressed. 
The light-cone momentum fraction 
   $z=(E+p_\|)_h/(E+p)_q$, where $p_\|$ is the momentum of the formed hadron $h$ 
   along the direction of the quark $q$, 
is given by the string-fragmentation function
$$
   f(z) \sim \frac{1}{z} (1-z)^a \exp\left(-\frac{b m_\perp^2}{z}\right) 
\EQN eqn-string-fragfct $$
where $a$ and $b$ are free parameters. These parameters need to be adjusted
to bring the fragmentation into accordance with measured data, {\it e.g.} 
$a=0.11$ and $b=0.52$ GeV$^{-2}$ as determined in \cite{bib-OPAL-ZPC69-543} (for an
overview see \cite{bib-Schmelling-habil}).

\subsection{Cluster fragmentation}
Assuming a local compensation of colour based on the {\it pre-confinement}
property of perturbative QCD \cite{bib-Amati-Veneziano}, the remaining gluons
at the end of the parton shower evolution are split non-perturbatively into 
quark-antiquark pairs. Colour singlet clusters of typical mass of a couple 
of GeV are then formed from quark and antiquark of colour-connected splittings. 
These clusters decay directly into two hadrons unless they are either too 
heavy (relative to an adjustable parameter {\tt CLMAX}, default value $3.35$~GeV), 
when they decay into 
two clusters, or too light, in which case a cluster decays into a single 
hadron, requiring a small rearrangement of energy and momentum with
neighbouring clusters. The decay of a cluster into two hadrons is assumed to
be isotropic in the rest frame of the cluster except if a perturbative-formed 
quark is involved. A decay channel is chosen based on the phase-space probability, 
the density of states, and the spin degeneracy of the hadrons. Cluster 
fragmentation has a compact description with few parameters, due to the
phase-space dominance in the hadron formation.

\section{Experimental studies}
A great wealth of measurements of $e^+e^-$ fragmentation into identified 
particles exists. A collection of references, where data on the fragmentation 
into identified particles can be found, is given for Tab.~37.1 of \cite{bib-RPP00}.   
As representatives
of all the data, Fig.~\ref{fig-pi-K-p-spectra} shows fragmentation functions
as the scaled momentum spectra of charged particles at several c.m.\ energies, 
$\sqrt{s}$.  In Fig.~\ref{fig-spectra-at-91GeV} $x_p$ spectra at 
$\sqrt{s}=91$ GeV are shown for all charged particles, for several identified 
charged and neutral particles.
Heavy flavour particles are dealt with separately in 
Sect.~\ref{sec-heavy-quark-frag}.
\figure 
\centerline{\epsfysize=18cm \epsfbox{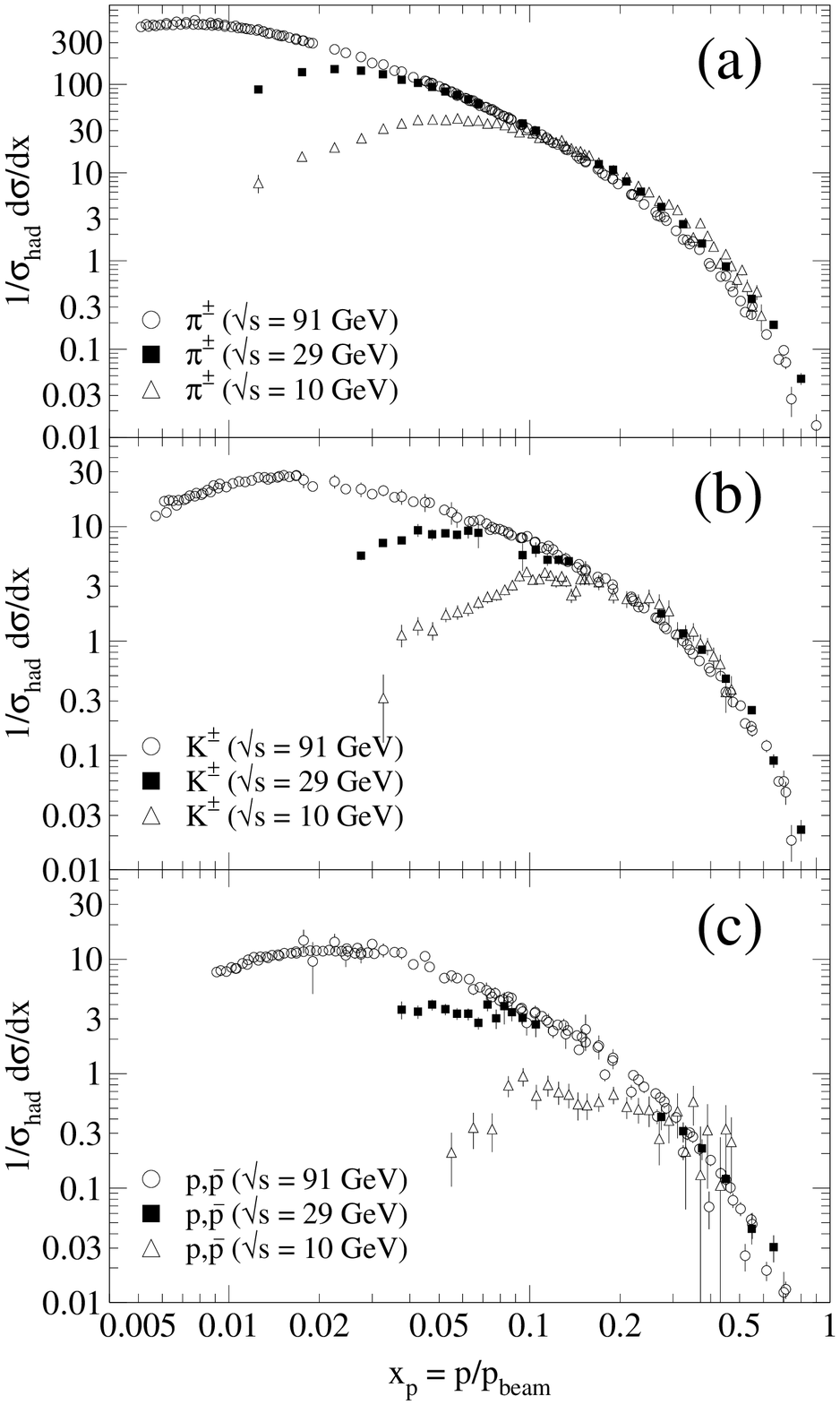}}
\Caption
         \label{fig-pi-K-p-spectra}
         Scaled momentum, $x_p \equiv 2p/\sqrt{s} = p/p_{\rm beam}$ spectra 
         of (a) $\pi^\pm$, (b) $K^\pm$, and (c) $p/\overline{p}$ at $\sqrt{s}=10$, 
         $29$, and $91$ GeV are shown \cite{
         bib-TPC-PRL61-1263,
         bib-ALEPH-PPE94-201,
         bib-ARGUS-ZPC44-547,
         bib-DELPHI-EPJC5-585,
         bib-OPAL-ZPC63-181,
         bib-SLD-PRD59-052001}.
\endCaption
\endfigure
\figure 
\centerline{\epsfysize=15cm \epsfbox{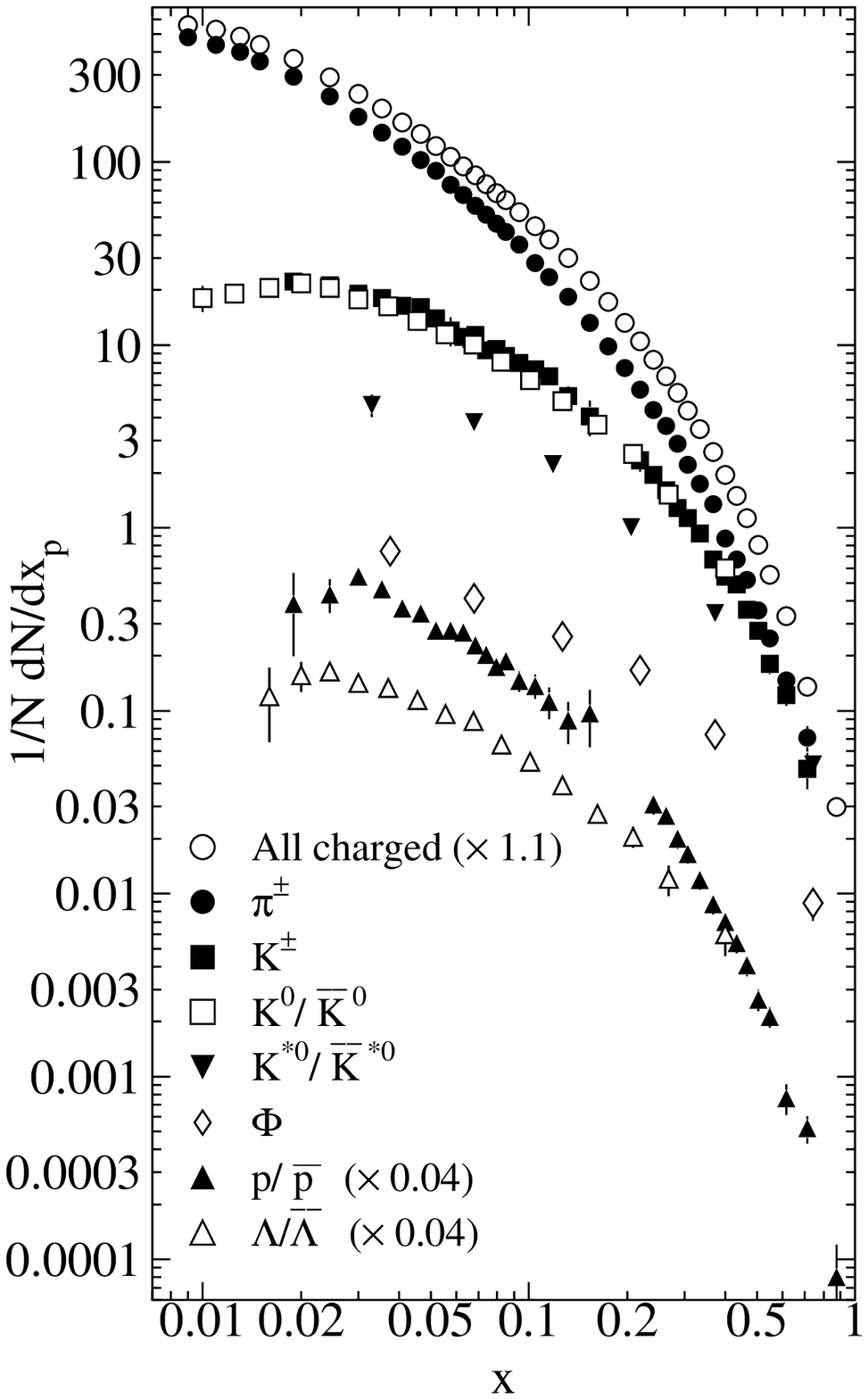}}
\Caption
         \label{fig-spectra-at-91GeV}
         Scaled momentum spectra $\sqrt{s}=91$ GeV for all charged particles, 
         for identified charged ($\pi^\pm$, $K^\pm$, $p/\overline{p}$) and for 
         identified neutral particles ($K_{\rm S}^0/\overline{K}_{\rm S}^0$, 
         $K^{*0}/\overline{K}^{*0}$, $\phi$, $\Lambda/\overline{\Lambda}$) are 
         shown \cite{bib-SLD-PRD59-052001}.
\endCaption
\endfigure

The measured fragmentation functions are solutions to the DGLAP equation \Ep{APDi}
but need to be parametrized at some initial scale $t_0$ (usually $2$ GeV$^2$ for
light quarks and gluons). A general parametrization is
\cite{bib-Kniehl-etal-NPB597-337}
$$
   D_{p\rightarrow h}(x,t_0) = N x^\alpha (1-x)^\beta \left(1+\frac{\gamma}{x}\right)
\EQN eqn-fragfct-param $$
where the normalization $N$, and the parameters $\alpha$, $\beta$, and $\gamma$ 
in general depend on the energy scale $t_0$ and also on the type of the parton, $p$,
and the hadron, $h$. Frequently the term involving $\gamma$
is left out \cite{Binn,
                  bib-Bourhis-etal-EPJC19-89, 
                  bib-Kniehl-etal-NPB582-514, 
                  bib-Binnewies-etal-PRD52-4947, 
                  bib-Binnewies-etal-ZPC65-471}. 
In the above quoted references the parameters of \Eqn{eqn-fragfct-param} 
are tabulated for many different combinations of partons and hadrons in 
$p\rightarrow h$. The parameters were obtained by fitting data for many 
different hadron types over a vast range of c.m.\ energies 
($\sqrt{s}\approx 5$ - $200$ GeV).

\section{Heavy quark fragmentation}
\label{sec-heavy-quark-frag}
It was recognized very early \cite{hardfrag} that
a heavy flavoured meson should retain a large fraction of the momentum
of the primordial heavy quark, and therefore its fragmentation function
should be much harder than that of a light hadron. In the limit
of a very heavy quark, one expects the fragmentation function
for a heavy quark to go into any heavy hadron to be peaked near 1.

When the heavy quark is produced at a momentum
much larger than its mass, one expects important perturbative effects,
enhanced by powers of the logarithm of the transverse momentum over the
heavy quark mass, to intervene and modify the shape of the fragmentation
function. In leading logarithmic order (\ie, including all powers
of $\as\log m_{\rm Q}/p_T$) the total (\ie, summed over all hadron types)
perturbative fragmentation function is simply obtained by solving the leading
evolution equation for fragmentation functions, \Eqn{APDi},
with the initial condition at a scale $\mu^2=m_{\rm Q}^2$ given by
$$
D_{\rm Q}(z,m_{\rm Q}^2)=\delta(1-z)\,,\quad D_{q}=0\,,\quad D_{\bar{q}}=0\,,
\quad D_{\bar{\rm Q}}=0\,,
\quad D_{ g}=0\,\quad 
\EQN D_x-lowest-order $$
where the notation $D_i(z)$ stands now for the probability to produce
a heavy quark ${\rm Q}$ from parton $i$ with a fraction $z$ of the parton momentum. 
If the scale $\mu^2$ is not too large, and one looks at relatively large
values of $z$, one can assume that the non-singlet evolution embodies
most of the physics. Following this assumption, for example,
the average value of $z$ at a scale $\mu^2$ is easily obtained by solving
the evolution equation in the moment representation 
(\cf\ \Eqn{scaviol})
$$
\langle z\rangle=\tilde{D}_{\rm Q}(2,\mu^2)=\left(\frac{\as(\mu^2)}{\as(m_{\rm Q}^2)}
\right)^{\frac{8 C_{F}}{11 C_{A} -2 n_{f}}}
\EQN mean_z $$
Several extensions of the leading logarithmic result have appeared
in the literature:
\begin{itemize}
\item
      Next-to-leading-log (NLL) order results for the perturbative heavy quark
      fragmentation function have been obtained in \cite{MeleNason}.
\item
      At large $z$, phase space for gluon radiation is suppressed. This
      exposes large perturbative corrections due to the incomplete
      cancellation of real gluon radiation and virtual gluon exchange
      (Sudakov effects), which should be resummed in order to get accurate
      results. A leading-log (LL) resummation formula has been obtained
      in ~\cite{MeleNason,DokshitzerKhoze}. Next-to-leading-log resummation has 
      been performed in \cite{CacCat}.
\item
      Fixed order calculations of the fragmentation function at order
      $\as^2$ in $e^+e^-$ annihilation have appeared in \cite{NasonOleari}.
      This result does not include terms of order $(\as\log s/m^2)^k$
      and $\as(\as\log s/m^2)^k$, but it does include correctly all terms
      up to the order $\as^2$, including terms without any logarithmic 
      enhancements.
\end{itemize}
Inclusion of non-perturbative effects in the calculation of
the heavy quark fragmentation function is done in practice by convolving
the perturbative result with a phenomenological non-perturbative
form.
Among the most popular parametrizations we have the following:
$$
\EQNdoublealign{
\mbox{Peterson \etal\ \cite{Peterson}:}&
D_{\rm np}(z)\propto & \frac{1}{z}
      \left(1-\frac{1}{z}-\frac{\epsilon}{1-z}\right)^{-2}\,,
\EQN eqn-Peterson \cr
\mbox{Kartvelishvili \etal\ \cite{Kartvelishvili}:}&
D_{\rm np}(z)\propto & z^\alpha (1-z)\,,
\EQN eqn-Kartvelishvili \cr
\mbox{Collins\&Spiller \cite{CollinsFR}:}&
D_{\rm np}(z)\propto & \left(\frac{1-z}{z}+\frac{(2-z)\epsilon_C}{1-z}\right) 
\times 
\cr
 && (1+z^2)\left(1-\frac{1}{z}-\frac{\epsilon_C}{1-z}\right)^{-2}
\EQN eqn-Collins \cr
}
$$
where $\epsilon$, $\alpha$, and $\epsilon_C$ are non-perturbative parameters,
depending upon the heavy hadron considered. In general,
the non-perturbative parameters do not have an absolute meaning.
They are fitted together with some model of hard radiation, which can be
either a shower Monte Carlo, a leading-log or NLL calculation
(which may or may not include Sudakov resummation), or a fixed order
calculation. In \cite{NasonOleari},
for example, the $\epsilon$ parameter for charm and bottom production
is fitted from the measured distributions of refs.~\cite{OPALCharm, ARGUSCharm} 
for charm, and of \cite{ALEPHbfrag1} for bottom.
If the leading-logarithmic approximation (LLA) is used for the
perturbative part, one finds
$\epsilon_c\approx 0.05$ and  $\epsilon_b\approx 0.006$; if a
second order calculation is used one finds $\epsilon_c\approx 0.035$ and
$\epsilon_b\approx 0.0033$; if a NLLO calculation is used instead one
finds  $\epsilon_c\approx 0.022$ and $\epsilon_b\approx 0.0023$.
The larger values found in the LL approximation are consistent with
what is obtained in the context of parton shower models \cite{Chrin},
as expected. The $\epsilon$ parameter for charm and bottom scales
roughly with the inverse square of the heavy flavour mass.
This behaviour can be justified by several arguments
\cite{hardfrag,JaffeRandall,NasonWebber}.
It can be used to relate the non-perturbative parts of the
fragmentation functions of charm and bottom quarks \cite{NasonOleari,
ColangeloNason, RandallRius}.

The bulk of the available fragmentation function data on charmed
mesons (excluding $J/\Psi(1S)$) is from measurements at $\sqrt{s}=10$ 
GeV.\footnote{This part on charm quark fragmentation including 
Fig.~\ref{fig-dfrag} and Table~\ref{RPP00-Tab.36.1}, which 
have been updated, are taken from the review of D.~Besson
contributed to C.~Caso \etal, \journal Eur.\ Phys.\ J.; C3, 1 (1998).} 
Shown in Fig.~\ref{fig-dfrag}
are the efficiency-corrected (but
not branching ratio corrected) CLEO \cite{bib-CLEO-PRD37-1719} and ARGUS
\cite{ARGUSCharm} inclusive cross-sections ($s\cdot {\cal B} d\sigma/dx_p$
in units of GeV$^2\,$nb, with $x_p=p/p_{\rm max}$) for the production of
pseudoscalar $D^0$ and vector $D^{*+}$ in $e^+e^-$ annihilation at
$\sqrt{s}\approx 10$ GeV. 
\figure 
\centerline{\epsfxsize=10cm \epsfbox{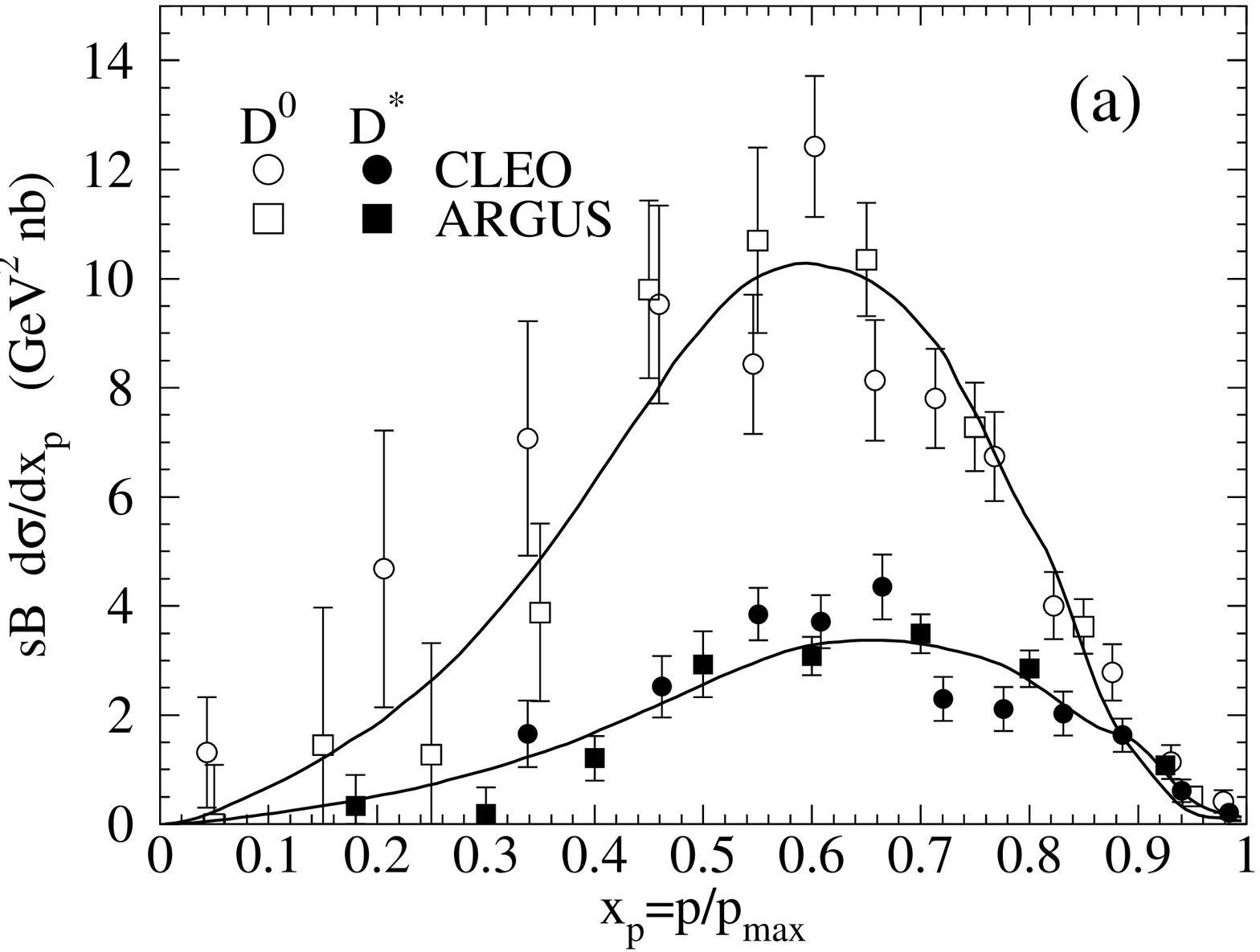}}
\Caption 
         \label{fig-dfrag}
         Efficiency-corrected inclusive cross-section measurements for the
         production of $D^0$ and $D^{*+}$ in $e^+e^-$ measurements at
         $\sqrt{s}\approx 10$ GeV. The variable $x_p$ approximates the
         Peterson variable $z$, but is not identical to it. 
\endCaption
\endfigure
For the $D^0$, ${\cal B}$ represents the product
branching fraction: $D^{*+} \rightarrow D^0 \pi^+$, $D^0 \rightarrow K^- \pi^+$.
These inclusive spectra have not been corrected for cascades from higher states, nor
for radiative effects. Note that since the momentum spectra are sensitive to
   QED and QCD 
radiative corrections, 
charm spectra at $\sqrt{s}=10$ GeV cannot be compared directly with spectra at
higher c.m.\ energies, and must be appropriately evolved.

   Tuning $\epsilon$ of the function \Ep{eqn-Peterson} in the JETSET 
   Monte Carlo generator \cite{bib-JETSET} using the parameter set of
   \cite{bib-OPAL-ZPC69-543} and including radiative corrections 
   to describe the combined CLEO and ARGUS $D^0$ and $D^{*+}$ data gives
   $\epsilon_c=0.043 \pm 0.004$ ($\chi^2/$d.o.f.$=50.5/41$);
   this is indicated in the solid curves. 
Measurements of the fragmentation functions for a variety of particles has 
allowed comparisons between mesons and baryons, and particles of different 
spin structure, as shown in Table \ref{RPP00-Tab.36.1}. 
   The $\epsilon$ values listed in this table were obtained from directly 
fitting the Peterson function \Ep{eqn-Peterson} to the measured differential 
cross-section $\sigma\cdot{\cal B}d\sigma/d x_p$.
\table 
\caption{
         \label{RPP00-Tab.36.1}
         The Peterson momentum hardness parameter $\epsilon$ as obtained
         from fits of \Eqn{eqn-Peterson} to $s\cdot{\cal B}d\sigma/dx_p$ of 
         $e^+e^- \rightarrow {\rm particle} + X$ measurements
         at $\sqrt{s}\approx 10$ GeV.
}
\centerline{\vbox{\halign{#\  \hfil & #\  \hfil & \hfil #\  \hfil & \hfil #\  \hfil \cr
\hline\hline
Particle        & $L$ &  $\epsilon$                   & Reference                      \cr
\hline
$D^0$           &  0  &  $0.260 \pm 0.024$            & \cite{bib-CLEO-PRD37-1719,
                                                              bib-ARGUS-PR276-223}     \cr
$D^+$           &  0  &  $0.156 \pm 0.022$            & \cite{bib-CLEO-PRD37-1719,
                                                              bib-ARGUS-PR276-223}     \cr
$D^{*+}$        &  0  &  $0.198 \pm 0.022$            & \cite{bib-CLEO-PRD37-1719,
                                                              bib-ARGUS-PR276-223}     \cr
$D_s$           &  0  &  $0.10 \pm 0.02$              & \cite{bib-CLEO-PRD62-072003,
                                                              bib-ARGUS-PR276-223}     \cr
$D_s^*$         &  0  &  $0.057 \pm 0.008$            & \cite{bib-CLEO-PRD62-072003,
                                                              bib-ARGUS-PR276-223}     \cr
\hline
$D_1(2420)^0$   &  1  &  $0.015 \pm 0.004$            & \cite{bib-CLEO-PLB331-236,
                                                              bib-ARGUS-PR276-223}     \cr
$D_2^*(2460)^0$ &  1  &  $0.039 \pm 0.013$            & \cite{bib-CLEO-PLB331-236,
                                                              bib-ARGUS-PR276-223}     \cr
$D_1(2420)^+$   &  1  &  $0.013 \pm 0.006$            & \cite{bib-CLEO-PLB341-435,
                                                              bib-ARGUS-PR276-223}     \cr
$D_2^*(2460)^+$ &  1  &  $0.023 \pm 0.009$            & \cite{bib-CLEO-PLB341-435,
                                                              bib-ARGUS-PR276-223}     \cr
$D_{s1}(2536)^+$&  1  &  $0.018 \pm 0.008$            & \cite{bib-CLEO-PLB303-377,
                                                              bib-ARGUS-PR276-223}     \cr
$D_{sJ}(2573)^+$&  1  &  $0.027 ^{+0.043}_{-0.015}$   & \cite{bib-ARGUS-ZPC69-405}     \cr
\hline\hline
$\Lambda_c^+$   &  0  &  $0.267\pm 0.038$             & \cite{bib-CLEO-PRD43-3599,
                                                              bib-ARGUS-PLB207-109}    \cr
$\Xi_c^{+,0}$   &  0  &  $0.23 \pm 0.05$              & \cite{bib-ARGUS-PLB247-121,
                                                              bib-CLEO-PLB373-261}     \cr
$\Xi_{c}'^{+,0}$&  0  &  $0.20 ^{+0.24}_{-0.11}$      & \cite{bib-CLEO-PRL82-492}      \cr
$\Sigma_c(2455)$&  0  &  $0.28 \pm 0.05$              & \cite{bib-ARGUS-PLB211-489,
                                                              bib-CLEO-PRL62-1240}     \cr
$\Sigma_c(2520)$&  0  &  $0.30 ^{+0.10}_{-0.07}$      & \cite{bib-CLEO-PRL78-2304}     \cr
$\Xi_c(2645)^{+}$& 0  &  $0.24 ^{+0.22}_{-0.10}$      & \cite{bib-CLEO-PRL77-810}      \cr
$\Xi_c(2645)^{0}$& 0  &  $0.22 ^{+0.15}_{-0.08}$      & \cite{bib-CLEO-PRL75-4364}     \cr
\hline
$\Lambda_{c}(2593)^+$ 
                &  1  &  $0.058 \pm 0.022$            & \cite{bib-CLEO-PRL74-3331,
                                                              bib-ARGUS-PLB402-207}    \cr
$\Lambda_{c}(2625)^+$ 
                &  1  &  $0.053 \pm 0.014$            & \cite{bib-CLEO-PRL74-3331,
                                                              bib-ARGUS-PLB317-227}    \cr
$\Xi_{c}(2815)$ &  1  &  $0.07 ^{+0.03}_{-0.02}$      & \cite{bib-CLEO-PRL83-3390}     \cr
\hline\hline
}}}
\endtable

We note from Table \ref{RPP00-Tab.36.1} that the mass dependence of $\epsilon$ is
less marked than the dependence on the orbital angular momentum structure of the charmed
hadron being measured. Orbitally excited $L=1$ charmed hadrons ($D_J$, $D_{sJ}$, and
$\Lambda_{c}(2593)$, $\Lambda_{c}(2625)$) show consistently harder spectra (\ie, smaller 
values of $\epsilon$)
than the $L=0$ ground states, whereas the data for the ground state charmed baryons
$\Lambda_c$ and $\Xi_c$ show agreement with the lighter (by $\approx 400$-$600$ MeV)
ground-state $D$ and $D_s$ charmed mesons. To some extent, the harder spectra of
$L=1$ hadrons can be attributed to the fact that all the $L=1$ charmed hadrons 
will eventually decay into $L=0$ hadrons.

Experimental studies of the fragmentation function for $b$ quarks
have been performed at LEP and SLD
\cite{ALEPHbfrag1,L3bfrag,OPALbfrag,SLDprecise}.
The results are shown in Fig.~\ref{fig-bfrag}.
\figure 
\centerline{\epsfxsize=10cm \epsfbox{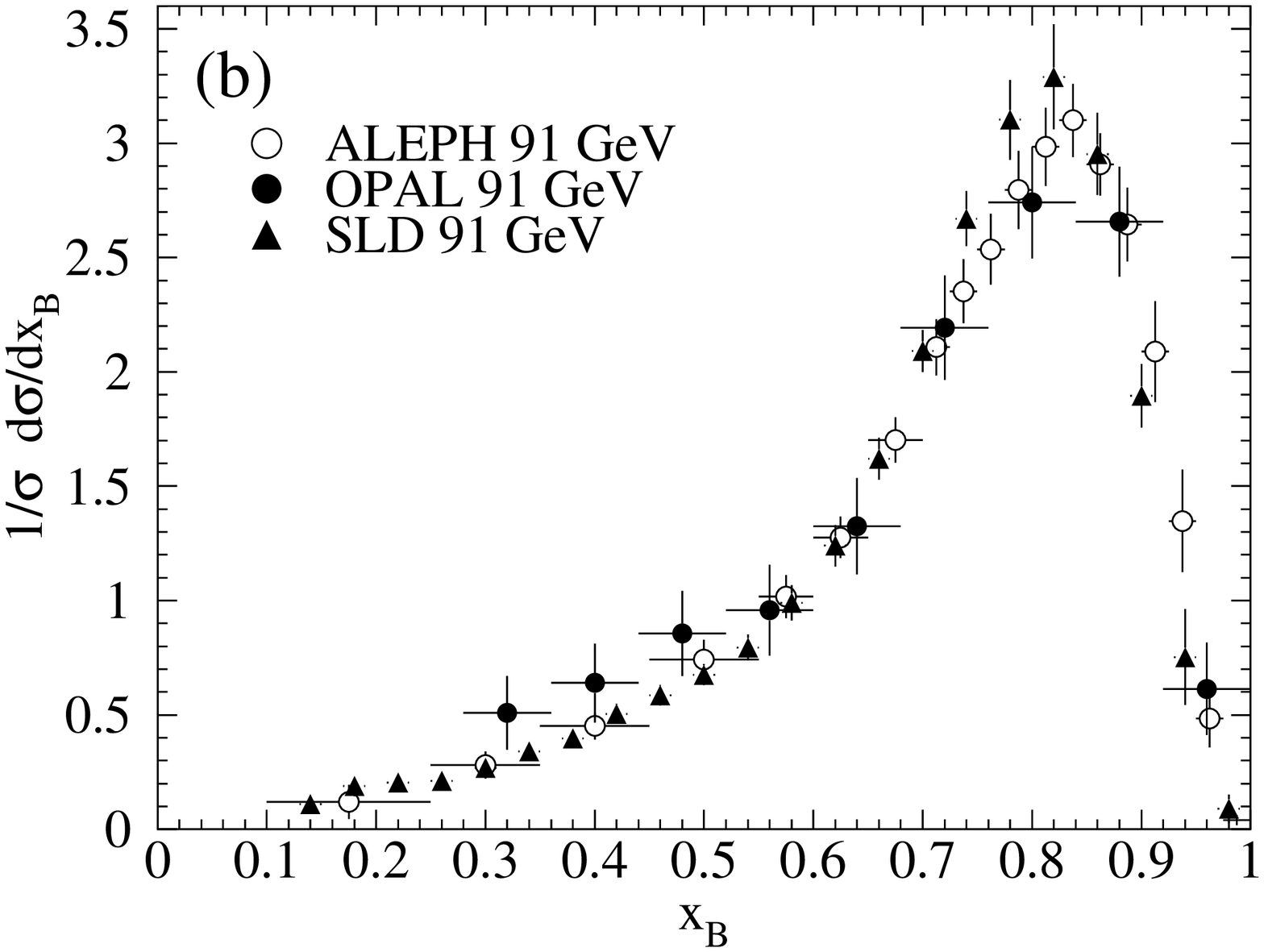}}
\Caption
         \label{fig-bfrag}
         Measured $e^+e^-$ fragmentation function of $b$ quarks into 
         $B$ hadrons \cite{OPALbfrag,
                           SLDprecise,
                           ALEPHbfrag}.
\endCaption
\endfigure
$B$ hadrons are commonly identified by their semileptonic decays or 
by their decay vertex reconstructed from the charged particles 
emerging from the $B$ decay.
The most recent results of refs.~\cite{ALEPHbfrag} and \cite{SLDprecise}
are in agreement.
Both studies fit the $B$ spectrum using
a Monte Carlo shower model supplemented with non-perturbative 
fragmentation functions to fit their data.

The experiments measure primarily
the spectrum of $B$ meson. This defines a fragmentation function
which includes the effect of the decay of higher mass excitations,
like the $B^*$ and $B^{**}$. There is an ambiguity in the definition
of the fragmentation function, which has to do with what is
considered to be a final state hadron. One may prefer to distinguish
$B$ meson produced directly from those arising from decays of $B^*$
and $B^{**}$, and define
$$
D^{(d)}_{B|B^*|B^{**}}(z)=D^{(d)}_{B}(z)+D^{(d)}_{B^*}(z)+D^{(d)}_{B^{**}}(z)
\EQN D_B|B*|B** $$
where the notation $D^{(d)}$ stands for directly produced.
The fragmentation function which is directly measured from the $B$ mesons,
irrespective of their origin, is instead
given by
$$
\EQNdoublealign{
D_B(x) &=& D^{(d)}_{B}(x)+\int F_{B^* \to B}(y) D^{(d)}_{B^*}(z) \delta(x-yz) dy dz
\cr
&&
\phantom{D^{(d)}_B(x)}
+\int F_{B^{**} \to B}(y) D^{(d)}_{B^{**}}(z) \delta(x-yz) dy dz\;.
\EQN conv_integral %
}
$$
where (for example) $F_{B^{**} \to B}(y)$ stands for the probability
for a fast $B^{**}$ to decay to  $B$ with a fraction $y$ of its energy.
Thus, in order to extract $D^{(d)}_{B|B^*|B^{**}}$ from $D_B$ a correction
factor should be computed, which gives a harder fragmentation
function. Thus, for example, ref.~\cite{SLDprecise} obtains
the value $\langle x_B \rangle=0.714\pm 0.005 ({\rm stat.}) \pm
0.007 ({\rm syst.}) \pm 0.002 ({\rm model})$; postulating a $B^*$
fraction of 0.75, one gets $\langle x_{B|B^*} \rangle=0.718$;
postulating a $B^{**}$ fraction of 0.25 yields
$\langle x_{B|B^{**}} \rangle=0.728$.

Besides degrading the fragmentation function by gluon radiation,
QCD evolution can also generate soft heavy quarks, increasing in 
the small $x$ region as $s$ increases. This effect has been studied
both theoretically and experimentally.
One important issue is to understand how often $b\bar{b}$ or
$c\bar{c}$ pairs are produced indirectly, via a gluon splitting mechanism.
Several theoretical studies are available on this topic
\cite{Mueller86,Mangano92a,Seymour95,MillerSeymour}.               
Experimental studies on charm production via gluon splitting
have been presented in refs. \cite{OPALgcc,ALEPHgcc},
and measurements of $g\to b\bar{b}$ have been given in
\cite{DELPHIgbb,ALEPHgbb,SLDgbb}.
The reported values are given in Table~\ref{tab:gqq}.
\table  
\caption{
         \label{tab:gqq}
         Fraction of events containing $g \to c\bar{c}$ and $g \to b\bar{b}$
         subprocesses in $Z$ decays, as measured by the various collaborations,
         compared with theoretical predictions.
         The central/lower/upper values for the theoretical predictions are 
         obtained with $m_c=(1.5 \pm 0.3)$ and $m_b=(4.75\pm 0.25)\,$GeV.
}
\centerline{\vbox{\halign{#\  \hfil & #\  \hfil & #\hfil \cr
\hline\hline
       & $\bar{n}_{g\to c\bar{c}}$ (\%)              & $\bar{n}_{g\to b\bar{b}}$  (\%)                \cr \hline
OPAL   & \cite{OPALgcc}  $  3.20 \pm 0.21 \pm 0.38 $ &                                                \cr 
ALEPH  & \cite{ALEPHgcc} $  2.65 \pm 0.74 \pm 0.51 $ & \cite{ALEPHgbb}  $ 0.277 \pm 0.042 \pm 0.057 $ \cr 
DELPHI &                                             & \cite{DELPHIgbb} $ 0.21  \pm 0.11  \pm 0.09  $ \cr 
SLD    &                                             & \cite{SLDgbb}    $ 0.307 \pm 0.071 \pm 0.066 $ \cr \hline
Theory \cite{Seymour95} &                            &                                                \cr
$\lambdamsb=150\,$MeV & \addsp{$ 1.35{+0.48 \atop -0.30} $} & $ 0.20 \pm 0.02$                        \cr
$\lambdamsb=300\,$MeV & \addsp{$ 1.85{+0.69 \atop -0.44} $} & $ 0.26 \pm 0.03$                        \cr 
\hline\hline
}}}
\endtable
In ref.~\cite{Seymour95} an explicit calculation
of these quantities has been performed.
Using these results charm and bottom
multiplicities for different values of the masses and of $\lambdamsb$
were computed in \cite{Buras2}.
They are reported in Table~\ref{tab:gqq}.
As can be seen, charm measurements are somewhat in excess of the
predictions. 
The averaged experimental result for charm,
$(3.10\pm 0.34)$\%, is 1 - 2 standard deviations outside the range
of the theoretical prediction,
preferring lower values of the quark mass and/or a larger value of $\lambdamsb$.
However, higher-order corrections may well be substantial at the charm quark mass 
scale.  Better agreement is achieved for bottom.

As reported in ref.~\cite{Seymour95},
Monte Carlo models are in qualitative agreement with these results,
although the spread of the values they obtain is somewhat larger
than the theoretical error estimated by the direct calculation.
In particular, for charm one finds that while HERWIG \cite{Marchesini92}
and JETSET \cite{bib-JETSET} 
agree quite
well with the theoretical calculation, ARIADNE \cite{Lonnblad92} 
is higher by roughly a factor of 2, and thus is in better agreement 
with data. 
For bottom, agreement between theory, models and data is adequate. For
a detailed discussion see ref.~\cite{LEP2YB}.

The discrepancy with the charm prediction may be due to experimental
cuts forcing the final state configuration to be more 3-jet like,
which increases the charm multiplicity. Calculations that take this
possibility into account are given in \cite{MillerSeymour}.

\section*{Acknowledgements}
The work of P.N.\ and B.R.W.\ was supported in part by the EU Fourth
Framework Programme `Training and Mobility of Researchers', Network
`Quantum Chromodynamics and the Deep Structure of Elementary Particles',
contract FMRX-CT98-0194 (DG 12 - MIHT).


\begin{thebibliography}{999}
\bibitem{bib-RPP00}
    {\sc Particle Data Group}, D.E.~Groom \etal, {\em ``Review of Particle Physics''},
    \journal Eur.\ Phys.\ J.; C15, 1 (2000).
\bibitem{DGLAP}
L.N.~Lipatov, \journal Sov.\ J.\ Nucl.\ Phys.; 20, 95 (1975);
V.N.~Gribov and L.N.~Lipatov, \journal Sov.\ J.\ Nucl.\ Phys.; 15, 438 (1972);
G.~Altarelli and G.~Parisi, \journal Nucl.\ Phys.; B126, 298 (1977);
Yu.L.~Dokshitzer, \journal Sov.\ Phys.\ JETP; 46, 641 (1977).
%
\bibitem{bib-Furmanski-PLB97-437}
W.~Furmanski and R.~Petronzio, preprint TH.2933--CERN (1980), 
   \journal Phys.\ Lett.; 97B, 437 (1980).
%
\bibitem{NW}
       P.~Nason and B.R.~Webber, \journal Nucl.\ Phys.; B421, 473 (1994);
       erratum \journal \ibid; B480, 755 (1996). 
%
\bibitem{bib-ESW-book}
 R.K.~Ellis, J.~Stirling, and B.R.~Webber: 
       {\it QCD and Collider Physics}\/, Cambridge University Press, Cambridge (1996).
%
\bibitem{bib-TASSO-PLB114-65}
{\sc tasso} collaboration: R.~Brandelik \etal, \journal Phys.\ Lett.; B114, 65 (1982).
%
\bibitem{bib-TASSO-ZPC47-187}
{\sc tasso} collaboration: W.~Braunschweig \etal, \journal Z.\ Phys.; C47, 187 (1990).
%
\bibitem{bib-HRS-PRD31-1}
{\sc hrs} collaboration: D.Bender \etal, \journal Phys.\ Rev.; D31, 1 (1984).
%
\bibitem{bib-MARKII-PRD37-1}
{\sc mark ii} collaboration: A.~Petersen \etal, \journal Phys.\ Rev.; D37, 1 (1988).
%
\bibitem{bib-TPC-PRL61-1263}
{\sc tpc} collaboration: H.~Aihara \etal, \journal Phys.\ Rev.\ Lett.; 61, 1263 (1988).
%
\bibitem{bib-AMY-PRD41-2675}
{\sc amy} collaboration: Y.K.~Li \etal, \journal Phys.\ Rev.; D41, 2675 (1990).
%
\bibitem{bib-ALEPH-PREP294-1}
{\sc aleph} collaboration: E.~Barate \etal, \journal Phys.\ Rep.; 294, 1 (1998).
%
\bibitem{bib-DELPHI-EPJC6-19}
{\sc delphi} collaboration: P.~Abreu \etal, \journal Eur.\ Phys.\ J.; C6, 19 (1999).
%
\bibitem{bib-L3-PLB259-199}
{\sc l3} collaboration: B.~Adeva \etal, \journal Phys.\ Lett.; B259, 199 (1991).
%
\bibitem{bib-OPAL-EPJC7-369}
{\sc opal} collaboration: K.~Ackerstaff \etal, \journal Eur.\ Phys.\ J.; C7, 369  (1998).
%
\bibitem{bib-MARKII-PRL64-1334}
{\sc mark ii} collaboration: G.S.~Abrams \etal, \journal Phys.\ Rev.\ Lett.; 64, 1334 (1990).
%
\bibitem{bib-ALEPH-ZPC73-409}
{\sc aleph} collaboration: D.~Buskulic \etal, \journal Z.\ Phys.; C73, 409 (1997).
%
\bibitem{bib-OPAL-ZPC72-191}
{\sc opal} collaboration: R.~Akers \etal, \journal Z.\ Phys.; C72, 191 (1996).
%
\bibitem{bib-OPAL-ZPC75-193}
{\sc opal} collaboration: K.~Ackerstaff \etal, \journal Z.\ Phys.; C75, 193 (1997).
%
\bibitem{bib-OPAL-EPJC16-185}
{\sc opal} collaboration: G.~Abbiendi \etal, \journal Eur.\ Phys.\ J.; C16, 185 (2000).
%
\bibitem{bib-ALEPH-EPJC17-1}
{\sc aleph} collaboration: R.~Barate \etal, \journal Eur.\ Phys.\ J.; C17, 1 (2000).
%
\bibitem{bib-DELPHI-PLB398-194}
{\sc delphi} collaboration: P.~Abreu \etal, \journal Phys.\ Lett.; B398, 194 (1997).
%
\bibitem{bib-OPAL-EPJC11-217}
{\sc opal} collaboration: G.~Abbiendi \etal, \journal Eur.\ Phys.\ J.; C11, 217 (1999).
%
\bibitem{bib-OPAL-ZPC68-203}
{\sc opal} collaboration: R.~Akers \etal, \journal Z.\ Phys.; C86, 203 (1995). 
%
\bibitem{bib-OPAL-ZPC68-179}
{\sc opal} collaboration: R.~Akers \etal, \journal Z.\ Phys.; C68, 179 (1995). 
%
\bibitem{BalBra}
       I.I.~Balitsky and V.M.~Braun, \journal Nucl.\ Phys.; B361, 93 (1991).
\bibitem{DasWeb}
       M.~Dasgupta and B.R.~Webber, \journal Nucl.\ Phys.; B484, 247 (1997).
\bibitem{bib-DELPHI-PLB311-408}
{\sc delphi} collaboration: P.~Abreu \etal, \journal Phys.\ Lett.; B311, 408 (1993); 
W.~de Boer and T.~Ku{\ss}maul, IEKP-KA/93-8, \hepph{9309280}.
%
\bibitem{bib-ALEPH-PLB357-487}
{\sc aleph} collaboration: D.~Barate \etal, \journal Phys.\ Lett.; B357, 487 (1995),
erratum \journal \ibid; B364, 247 (1995).
%
\bibitem{bib-DELPHI-EPJC13-573}
{\sc delphi} collaboration: P.~Abreu \etal, \journal Eur.\ Phys.\ J.; C13, 573 (2000).
%
\bibitem{bib-Kniehl-PRL85-5288}
B.A.~Kniehl, G.~Kramer, and B.~P{\"o}tter, \journal Phys.\ Rev.\ Lett.; 85, 5288 (2001).
%
\bibitem{bib-colour-coherence}
B.I.~Ermolayev and V.S.~Fadin, \journal JETP Lett.; 33, 285 (1981);
A.H.~Mueller, \journal Phys.\ Lett.; B104, 161 (1981).
%
\bibitem{book}
       Yu.L.~Dokshitzer, V.A.~Khoze, A.H.~Mueller and S.I.~Troyan,
       {\it Basics of Perturbative QCD}\/, Editions Fronti\`eres, Paris (1991).
\bibitem{Mueller}
       A.H.~Mueller,  \journal Nucl.\ Phys.; B213, 85 (1983); erratum quoted in
       \journal \ibid; B241, 141 (1984); \journal \ibid; B228, 351 (1983).
\bibitem{bib-LENA-ZPC9-1}
{\sc lena} collaboration: B.~Niczporuk \etal, \journal Z.\ Phys.; C9, 1 (1981).
%
\bibitem{bib-CLEO-PRL49-357}
{\sc cleo} collaboration: M.S.~Alam \etal, \journal Phys.\ Rev.\ Lett.; 49, 357 (1982). 
%
\bibitem{bib-JADE-ZPC20-187}
{\sc jade} collaboration: W.~Bartel \etal, \journal Z.\ Phys.; C20, 187 (1983). 
%
\bibitem{bib-HRS-PRD34-3304}
{\sc hrs} collaboration: M.~Derrick \etal, \journal Phys.\ Rev.; D34, 3304 (1986). 
%
\bibitem{bib-TPC-PL134B-299}
{\sc tpc/two-gamma} collaboration: H.~Aihara \etal, \journal Phys.\ Lett.; 134B, 299 (1987). 
%
\bibitem{bib-AMY-PRD42-737}
{\sc amy} collaboration: H.W.~Zheng \etal, \journal Phys.\ Rev.; D42, 737 (1990). 
%
\bibitem{bib-ALEPH-98-025}
{\sc aleph} collaboration: 
``QCD studies with $e^+e^-$ annihilation data from $130$ to $183$~GeV'',
  contributed paper to ICHEP98 {\bf \#945} (1998).
%
\bibitem{bib-DELPHI-PLB372-172}
{\sc delphi} collaboration: P.~Abreu \etal, \journal Phys.\ Lett.; B372, 172 (1996). 
%
\bibitem{bib-DELPHI-PLB416-233}
{\sc delphi} collaboration: P.~Abreu \etal, \journal Phys.\ Lett.; B416, 233 (1998). 
%
\bibitem{bib-L3-PLB371-137}
{\sc l3} collaboration: M.~Acciarri \etal, \journal Phys.\ Lett.; B371, 137 (1996). 
%
\bibitem{bib-L3-PLB404-390}
{\sc l3} collaboration: M.~Acciarri \etal, \journal Phys.\ Lett.; B404, 390 (1997). 
%
\bibitem{bib-L3-2304}
{\sc l3} collaboration:
``Preliminary Cross Section Measurements from the {\sc l3} experiment at
  $\sqrt{s} = 189$~GeV'', contributed paper to ICHEP98 {\bf \#484}
  (1998).
%
\bibitem{bib-L3-98-148}
{\sc l3} collaboration: M.~Acciarri \etal, \journal Phys.\ Lett.; B444, 569 (1998). 
%
\bibitem{bib-Dremin}
I.M.~Dremin, \journal JETP Lett.; 68, 559 (1998), \hepph{9808481}.
%
\bibitem{bib-Dremin-Nechitailo}
I.M.~Dremin and V.A.~Nechitailo, \journal Mod.\ Phys.\ Lett.; A9, 1471 (1994),
  \hepex{9406002}.
%
\bibitem{bib-Dremin-Gary-PLB459-341}
I.M.~Dremin and J.W.~Gary, \journal Phys.\ Lett.; B459, 341 (1999),
  \hepph{9905477}.
%
\bibitem{FongWeb}
       C.P.~Fong and B.R.~Webber, \journal Nucl.\ Phys.; B355, 54 (1992).
\bibitem{bib-TOPAZ-PLB345-335}
{\sc topaz} collaboration: R.~Itoh \etal, \journal Phys.\ Lett.; B345, 335 (1995). 
%
\bibitem{bib-DELPHI-ZPC73-11}
{\sc delphi} collaboration: P.~Abreu \etal, \journal Z.\ Phys.; C73, 11 (1996). 
%
\bibitem{bib-DELPHI-ZPC73-229}
{\sc delphi} collaboration: P.~Abreu \etal, \journal Z.\ Phys.; C73, 229 (1997). 
%
\bibitem{bib-TPC-LBL-23737}
{\sc tpc/two-gamma} collaboration: H.~Aihara \etal, 
``Charged Hadron Production in $e^+e^-$ Annihilation at $\sqrt{s}=29$~GeV'',
  LBL 23737.
%
\bibitem{bib-ALEPH-ZPC55-209}
{\sc aleph} collaboration: D.~Buskulic \etal, \journal Z.\ Phys.; C55, 209 (1992). 
%
\bibitem{bib-DELPHI-PLB275-231}
{\sc delphi} collaboration: P.~Abreu \etal, \journal Phys.\ Lett.; B275, 231 (1992). 
%
\bibitem{bib-DELPHI-98-83}
{\sc delphi} collaboration: P.~Abreu \etal, \journal Phys.\ Lett.; B459, 397 (1999). 
%
\bibitem{bib-OPAL-PLB247-617}
{\sc opal} collaboration: M.Z.~Akrawy \etal, \journal Phys.\ Lett.; B247, 617 (1990). 
%
\bibitem{CatTre}
       S.~Catani and L.~Trentadue, \journal Nucl.\ Phys.; B327, 353 (1989); 
       \journal \ibid; B353, 183 (1991).
\bibitem{CacCat}
       M.~Cacciari and S.~Catani, CERN-TH/2001-174, UPRF-2001-11, \hepph{0107138}.
\bibitem{CFP}
       G.~Curci, W.~Furmanski, and R.~Petronzio, \journal Nucl.\ Phys.; B175, 27 (1980);
       E.G.~Floratos, C.~Kounnas, and R.~Lacaze, \journal Nucl.\ Phys.; B192, 417 (1981).
\bibitem{AEMP}
       G.~Altarelli, R.K.~Ellis, G.~Martinelli and S-Y.~Pi,
       \journal Nucl.\ Phys.; B160, 301 (1979).
\bibitem{RvN}
       P.J.~Rijken and W.L.~van Neerven, \journal Nucl.\ Phys.; B487, 233 (1997).
\bibitem{Binn}
       J.~Binnewies, Hamburg University PhD Thesis, DESY 97-128,
       \hepph{9707269}.
\bibitem{bib-Artru-Mennessier}
X.~Artru and G.~Mennessier, \journal Nucl.\ Phys.; B70, 93 (1974). 
%
\bibitem{bib-Andersson-Gustafson}
B.~Andersson and G.~Gustafson, \journal Z.\ Phys.; C3, 223 (1980). 
%
\bibitem{bib-JETSET}
 T.~Sj{\"{o}}strand and M.~Bengtsson, \journal Comput.\ Phys.\ Commun.; 43, 367 (1987),
 T.~Sj{\"{o}}strand, \journal Comput.\ Phys.\ Commun.; 82, 74 (1994). 
\bibitem{bib-UCLA} 
S.~Chun and C.~Buchanan, \journal Phys.\ Rep.; 292, 239 (1998).
%
\bibitem{Marchesini92}
 G.~Marchesini, B.R.~Webber, G.~Abbiendi, I.G.~Knowles, M.H.~Seymour and
 L.~Stanco, \journal Comput.\ Phys.\ Commun.; 67, 465 (1992);
 G.~Corcella, I.G.~Knowles, G.~Marchesini, S.~Moretti, K.~Odagiri,
 P.~Richardson, M.H.~Seymour, and B.R.~Webber, \journal JHEP; 0101, 010 (2001).
%
\bibitem{bib-OPAL-ZPC69-543}
{\sc opal} collaboration: G.~Alexander \etal, \journal Z.\ Phys.; C69, 543 (1996). 
%
\bibitem{bib-Schmelling-habil}
M.~Schmelling, \journal Phys.\ Script.; 51, 683 (1995). 
%
\bibitem{bib-Amati-Veneziano}
D.~Amati and G.~Veneziano, \journal Phys.\ Lett.; B83, 87 (1979). 
%
\bibitem{bib-ALEPH-PPE94-201}
{\sc aleph} collaboration: D.~Buskulic \etal, \journal Z.\ Phys.; C66, 355 (1995).
%
\bibitem{bib-ARGUS-ZPC44-547}
{\sc argus} collaboration: H.~Albrecht \etal, \journal Z.\ Phys.; C44, 547 (1989).
%
\bibitem{bib-DELPHI-EPJC5-585}
{\sc delphi} collaboration: P.~Abreu \etal, \journal Eur.\ Phys.\ J.; C5, 585 (1998).
%
\bibitem{bib-OPAL-ZPC63-181}
{\sc opal} collaboration: R.~Akers \etal, \journal Z.\ Phys.; C, 181 (1994).
%
\bibitem{bib-SLD-PRD59-052001}
{\sc sld} collaboration: K.~Abe \etal, \journal Phys.\ Rev.; D59, 052001 (1999).
%
\bibitem{bib-Kniehl-etal-NPB597-337}
B.A.~Kniehl, G.~Kramer and B.~P{\"o}tter, \journal Nucl.\ Phys.; B597, 337 (2001). 
%
\bibitem{bib-Bourhis-etal-EPJC19-89}
L.~Bourhis, M.~Fontannaz, J.Ph.~Guillet, and M.~Werlen, 
\journal Eur.\ Phys.\ J.; C19, 89 (2001). 
%
\bibitem{bib-Kniehl-etal-NPB582-514}
B.A.~Kniehl, G.~Kramer, and B.~P{\"o}tter, \journal Nucl.\ Phys.; B582, 514 (2000). 
%
\bibitem{bib-Binnewies-etal-PRD52-4947}
J.~Binnewies, B.A.~Kniehl, and G.~Kramer, \journal Phys.\ Rev.; D52, 4947 (1995). 
%
\bibitem{bib-Binnewies-etal-ZPC65-471}
J.~Binnewies, B.A.~Kniehl, and G.~Kramer, \journal Z.\ Phys.; C65, 471 (1995). 
%
\bibitem{hardfrag}
    V.A.~Khoze, Ya.I.~Azimov, and L.L.~Frankfurt, Proceedings, Conference 
    on High energy physics, Tbilisi 1976;
    J.D.~Bjorken, \journal Phys.\ Rev.; D17, 171 (1978).
\bibitem{MeleNason}
    B.~Mele and P.~Nason, \journal Phys.\ Lett.; B245, 635 (1990), 
                          \journal Nucl.\ Phys.; B361, 626 (1991).
\bibitem{DokshitzerKhoze}
    Y.~Dokshitzer, V.A.~Khoze, and S.I.~Troyan, \journal Phys.\ Rev.; D53, 89 (1996).
%
\bibitem{NasonOleari}
    P.~Nason and C.~Oleari, \journal Phys.\ Lett.; B418, 199 (1998), \hepph{9709358},
    \journal Phys.\ Lett.; B447, 327 (1999), \hepph{9811206}, and
    \journal Nucl.\ Phys.; B565, 245 (2000), \hepph{9903541}.
\bibitem{Peterson}
    C.~Peterson \etal, \journal Phys.\ Rev.; D27, 105 (1983).
\bibitem{Kartvelishvili}
   V.G.~Kartvelishvili, A.K.~Likehoded, and V.A.~Petrov, \journal Phys.\ Lett.; B78, 615 (1978).
\bibitem{CollinsFR}
   P.~Collins and T.~Spiller, \journal J.\ Phys.; G11, 1289 (1985).
\bibitem{OPALCharm}
{\sc opal} collaboration: R.~Akers \etal, \journal Z. Phys.; C67, 27 (1995).
%
\bibitem{ARGUSCharm}
{\sc argus} collaboration: H.~Albrecht \etal, \journal Z.\ Phys.; C52, 353 (1991).
\bibitem{ALEPHbfrag1}
    {\sc aleph} collaboration: D.~Buskulic \etal, \journal Phys.\ Lett.; B357, 699 (1995).
\bibitem{Chrin}
    J.~Chrin, \journal Z.\ Phys.; C36, 163 (1987).
\bibitem{JaffeRandall}
    R.L.~Jaffe and L.~Randall, \journal Nucl.\ Phys.; B412, 79 (1994), \hepph{9306201}.
\bibitem{NasonWebber}
    P.~Nason and B.~Webber, \journal Phys.\ Lett.; B395, 355 (1997), \hepph{9612353}.
\bibitem{ColangeloNason}
    G.~Colangelo and P.~Nason, \journal Phys.\ Lett.; B285, 167 (1992).
\bibitem{RandallRius}
    L.~Randall and N.~Rius, \journal Nucl.\ Phys.; B441, 167 (1995), \hepph{9405217}.
\bibitem{bib-CLEO-PRD37-1719}
{\sc cleo} collaboration: D.~Bortoletto \etal, \journal Phys.\ Rev.; D37, 1719 (1988);
erratum \journal \ibid; 39, 1471 (1989).
%
\bibitem{bib-ARGUS-PR276-223}
{\sc argus} collaboration: H.~Albrecht \etal, \journal Phys.\ Rep.; 276, 223 (1996).
%
\bibitem{bib-CLEO-PRD62-072003} 
{\sc cleo} collaboration: R.A.~Briere \etal, \journal Phys.\ Rev.; D62, 072003 (2000).
%
\bibitem{bib-CLEO-PLB331-236} 
{\sc cleo} collaboration: P.~Avery \etal, \journal Phys.\ Lett.; B331, 236 (1994).
%
\bibitem{bib-CLEO-PLB341-435}
{\sc cleo} collaboration: T.~Bergfeld \etal, \journal Phys.\ Lett.; B340, 194 (1994).
%
\bibitem{bib-CLEO-PLB303-377}
{\sc cleo} collaboration: J.P.~Alexander \etal, \journal Phys.\ Lett.; B303, 377 (1993).
%
\bibitem{bib-ARGUS-ZPC69-405}
{\sc argus} collaboration: H.~Albrecht \etal, \journal Z.\ Phys.; C69, 405 (1996).
%
\bibitem{bib-CLEO-PRD43-3599}
{\sc cleo} collaboration: P.~Avery \etal, \journal Phys.\ Rev.; D43, 3599 (1993).
%
\bibitem{bib-ARGUS-PLB207-109}
{\sc argus} collaboration: H.~Albrecht \etal, \journal Phys.\ Lett.; B207, 109 (1988).
%
\bibitem{bib-ARGUS-PLB247-121}
{\sc argus} collaboration: H.~Albrecht \etal, \journal Phys.\ Lett.; B247, 121 (1990).
%
\bibitem{bib-CLEO-PLB373-261}
{\sc cleo} collaboration: K.W.~Edwards \etal, \journal Phys.\ Lett.; B373, 261 (1996).
%
\bibitem{bib-CLEO-PRL82-492}
{\sc cleo} collaboration: C.P.~Jessop \etal, \journal Phys.\ Rev.\ Lett.; 82, 492 (1999).
%
\bibitem{bib-ARGUS-PLB211-489}
{\sc argus} collaboration: H.~Albrecht \etal, \journal Phys.\ Lett.; B211, 489 (1988).
%
\bibitem{bib-CLEO-PRL62-1240}
{\sc cleo} collaboration: T.~Bowcock \etal, \journal Phys.\ Rev.\ Lett.; 62, 1240 (1989).
%
\bibitem{bib-CLEO-PRL78-2304}
{\sc cleo} collaboration: G.~Brandenberg \etal, \journal Phys.\ Rev.\ Lett.; 78, 2304 (1997).
%
\bibitem{bib-CLEO-PRL77-810}
{\sc cleo} collaboration: L.~Gibbons \etal, \journal Phys.\ Rev.\ Lett.; 77, 810 (1996).
%
\bibitem{bib-CLEO-PRL75-4364}
{\sc cleo} collaboration: P.~Avery \etal, \journal Phys.\ Rev.\ Lett.; 75, 4364 (1995).
%
\bibitem{bib-CLEO-PRL74-3331}
{\sc cleo} collaboration: K.W.~Edwards \etal, \journal Phys.\ Rev.\ Lett.; 74, 3331 (1995).
%
\bibitem{bib-ARGUS-PLB402-207}
{\sc argus} collaboration: H.~Albrecht \etal, \journal Phys.\ Lett.; B402, 207 (1997).
%
\bibitem{bib-ARGUS-PLB317-227}
{\sc argus} collaboration: H.~Albrecht \etal, \journal Phys.\ Lett.; B317, 227 (1993).
%
\bibitem{bib-CLEO-PRL83-3390}
{\sc cleo} collaboration: J.P.~Alexander \etal, \journal Phys.\ Rev.\ Lett.; 83, 3390 (1999).
%
\bibitem{L3bfrag}
    {\sc l3} collaboration: B.~Adeva \etal, \journal Phys.\ Lett.; B261, 177 (1991).
%
\bibitem{OPALbfrag}
    {\sc opal} collaboration: G.~Alexander \etal, \journal Phys.\ Lett.; B364, 93 (1995).
%
\bibitem{SLDprecise}
    {\sc sld} collaboration: K.~Abe \etal, SLAC-PUB-8504 (Jul 2000); 
    K.~Abe \etal, \journal Phys.\ Rev.\ Lett.; 84, 4300 (2000).
%
\bibitem{ALEPHbfrag}
    {\sc aleph} collaboration: A.~Heister \etal,
    \journal Phys.\ Lett.; B512, 30 (2001).
%
\bibitem{Mueller86}
  A.H.~Mueller and P.~Nason, \journal Nucl.\ Phys.; B266, 265 (1986).
%
\bibitem{Mangano92a}
  M.L.~Mangano and P.~Nason, \journal Phys.\ Lett.; B285, 160 (1992).
%
\bibitem{Seymour95}
  M.H.~Seymour, \journal Nucl.\ Phys.; B436, 163 (1995).
%
\bibitem{MillerSeymour}
D.J.~Miller and M.H.~Seymour, \journal Phys.\ Lett.; B435, 213 (1998), \hepph{9805414}.
%
\bibitem{OPALgcc}
 {\sc opal} collaboration: G.~Abbiendi \etal, \journal Eur.\ Phys.\ J.; C13, 1 (2000).
%
\bibitem{ALEPHgcc}
  {\sc aleph} collaboration: R.~Barate \etal, \journal Eur.\ Phys.\ J.; C16, 597 (2000).
%
\bibitem{DELPHIgbb}
{\sc delphi} collaboration:  P.~Abreu \etal, \journal Phys.\ Lett.; B405, 202 (1997).
%
\bibitem{ALEPHgbb}
{\sc aleph} collaboration: R.~Barate \etal,  \journal Phys.\ Lett.; B434, 437 (1998).
%
\bibitem{SLDgbb}
{\sc sld} collaboration: K.~Abe \etal, SLAC-PUB-8157, \hepex{9908028}.
%
\bibitem{Buras2}
    S.~Frixione, M.L.~Mangano, P.~Nason, and G.~Ridolfi: Heavy Quark Production,
    in  A.J.~Buras and M.~Lindner (eds.), {\it Heavy flavours II},
    World Scientific, Singapore (1998),
    \hepph{9702287}.
%
\bibitem{Lonnblad92}
  L.~L{\"{o}}nnblad, \journal Comput.\ Phys.\ Commun.; 71, 15 (1992).
%
\bibitem{LEP2YB}
A.~Ballestrero \etal, CERN-2000-09-B, \hepph{0006259}.
%
\end{thebibliography}
\end{document}